\documentclass[10pt,conference,letterpaper]{IEEEtran}

\pdfoutput=1

\newtheorem{definition}{Definition}[section]

\newtheorem{lemma}{Lemma}[section]

\newtheorem{theorem}{Theorem}[section]

\newtheorem{proposition}{Proposition}[section]

\newtheorem{example}{Example}[section]

\let\labelindent\relax
\usepackage[shortlabels]{enumitem}
\usepackage{balance}
\usepackage{url}
\usepackage{hyperref}
\usepackage{epsfig, amssymb, amsmath, bbm, multirow, graphicx}

\usepackage[caption=false,font=footnotesize]{subfig}
\usepackage[linesnumbered,ruled,noend]{algorithm2e}

\newcommand{\nop}[1]{}

\newcommand{\neighbor}[2][g]{\mathcal{N}_{#1}(#2)}
\newcommand{\pair}[2]{\langle #1 , #2 \rangle}

\begin{document}

\title{DDSL: Efficient Subgraph Listing on Distributed and Dynamic Graphs}


\author{\IEEEauthorblockN{1\textsuperscript{st} Xun Jian}
\IEEEauthorblockA{\textit{Department of Computer Science and Engineering} \\
\textit{Hong Kong University of Science and Technology}\\
Hong Kong, China \\
xjian@cse.ust.hk}
\and
\IEEEauthorblockN{2\textsuperscript{nd} Yue Wang}
\IEEEauthorblockA{\textit{Department of Computer Science and Engineering} \\
\textit{Hong Kong University of Science and Technology}\\
Hong Kong, China \\
ywangby@connect.ust.hk}
\and
\IEEEauthorblockN{3\textsuperscript{rd} Xiayu Lei}
\IEEEauthorblockA{\textit{Department of Computer Science and Engineering} \\
\textit{Hong Kong University of Science and Technology}\\
Hong Kong, China \\
xylei@cse.ust.hk}
\and
\IEEEauthorblockN{4\textsuperscript{th} Yanyan Shen}
\IEEEauthorblockA{\textit{Department of Computer Science and Engineering} \\
\textit{Shanghai Jiao Tong University}\\
Shanghai, China \\
shenyy@sjtu.edu.cn}
\and
\IEEEauthorblockN{5\textsuperscript{th} Lei Chen}
\IEEEauthorblockA{\textit{Department of Computer Science and Engineering} \\
\textit{Hong Kong University of Science and Technology}\\
Hong Kong, China \\
leichen@cse.ust.hk}
}


\maketitle
\begin{abstract}
Subgraph listing is a fundamental problem in graph theory and has wide applications in areas like sociology, chemistry, and social networks. Modern graphs can usually be large-scale as well as highly dynamic, which challenges the efficiency of existing subgraph listing algorithms. Recent works have shown the benefits of partitioning and processing big graphs in a distributed system, however, there is only few work targets subgraph listing on dynamic graphs in a distributed environment. In this paper, we propose an efficient approach, called \textit{Distributed and Dynamic Subgraph Listing} (DDSL), which can incrementally update the results instead of running from scratch. DDSL follows a general distributed join framework. In this framework, we use a Neighbor-Preserved storage for data graphs, which takes bounded extra space and supports dynamic updating. After that, we propose a comprehensive cost model to estimate the I/O cost of listing subgraphs. Then based on this cost model, we develop an algorithm to find the optimal join tree for a given pattern. To handle dynamic graphs, we propose an efficient left-deep join algorithm to incrementally update the join results. 
Extensive experiments are conducted on real-world datasets. The results show that DDSL outperforms existing methods in dealing with both static dynamic graphs in terms of the responding time.
\end{abstract}

\section{Introduction}
\label{sec:introduction}
In real-world applications, the underlying data can often be modeled as graphs. For example, the World Wide Web can be treated as a graph, where each single vertex represents a page, and each edge represents a hyper-link between two pages. Give another example, we can model a social network as a graph by treating users as vertices and friend relationships as edges.

In this paper, we study subgraph listing, one of the fundamental problems in graph theory. Given two undirected and unlabeled graphs $d$ and $p$, it requires to list all subgraphs of the \textit{data graph} $d$, which are isomorphic to the \textit{pattern graph} $p$. Such a subgraph is also called a match. Subgraph listing has wide applications in different areas. For example, in sociology, chemistry, telecommunication and bioinformatics, subgraph listing is used in comparing large graphs \cite{shervashidze2009efficient,prvzulj2007biological}. In activity networks, it is used to monitor potential terrorists by searching threat patterns \cite{cook2006mining}. It can also be adapted to track social network evolution \cite{Kairam:2012:LDO:2124295.2124374} and to identify properties that are useful in recommendation networks \cite{Leskovec:2006:PIR:2097044.2097097}.

\subsection{Motivation}
Traditional centralized solutions \cite{ullmann1976algorithm,chiba1985arboricity,kim2016dualsim} for subgraph listing try to enumerate subgraphs on a single machine. However, the result size can be exponential to the size of the data graph, which causes heavy I/O cost when we want to materialize the result on the disks. In this case, disk I/Os become the bottleneck of the whole algorithm, and centralized algorithms are thus not preferred in dealing with large graphs. To address this issue, other works \cite{Shao:2014:PSL:2588555.2588557,lai2015scalable,lai2016scalable,qiao2017subgraph,ammar2018distributed} focus on solving this problem in distributed environments. By partitioning and processing the data graph and the intermediate results (i.e., the partial matches) on the cloud, they reduce the I/O cost on each single machine and the total running time as well.

Another issue is the highly dynamic nature of modern graphs, where vertices and edges are added and deleted from time to time. For example, the number of monthly active users in Facebook increased to more than $1.94$ billion in 2017\footnote{https://www.statista.com/topics/751/facebook/} since it was founded in 2004. On average, in every second there are $4.7$ new users appear in this social network. In such a situation, when we want to monitor the matches of a certain pattern, it is not efficient to treat the updated graph as a new one and run the algorithm from scratch. Instead, only tracking the change of the result can be more efficient and thus is important.

Although existing distributed algorithms \cite{sun2012efficient,Shao:2014:PSL:2588555.2588557,lai2015scalable,lai2016scalable,qiao2017subgraph,ammar2018distributed} achieve good performance on static graphs, few of them targets exact subgraph listing on distributed and dynamic graphs. In practice, there are two major difficulties to handle dynamic graphs. Firstly, it requires efficient updating of the underlying data storage of $d$. In works \cite{lai2016scalable,qiao2017subgraph}, one of the preprocessing steps is to list all $k-$cliques for $k \leq k_0$ in $d$, which takes significant amount of time. When $d$ changes, whether the storage of cliques can be updated efficiently is unknown, so this could prevent the whole algorithm from efficiently handling dynamic graphs. Secondly, it also requires updating the result efficiently without doing too much redundant calculation, otherwise there would be no difference compared to running the original algorithm from scratch.


\subsection{Overview and Contributions}
In this paper, we propose \textit{Dynamic and Distributed Subgraph Listing} (DDSL), which attempts to solve the exact subgraph listing problem on distributed and dynamic graphs. DDSL can be fitted into general distributed data processing engines like MapReduce \cite{dean2008mapreduce}, Spark \cite{zaharia2010spark} and Dryad \cite{isard2007dryad}, and we describe it in MapReduce for the ease of presentation. In those systems, the I/O cost and the communication cost are often considered as the bottleneck of the whole algorithm \cite{lai2015scalable}, therefore, we conduct our analysis according to the amount of data involved in disk I/Os and communications at each step in our approach.

The whole approach of DDSL can be divided into two stages:
\begin{enumerate}
\item \textbf{Initial Calculation.} In this stage, DDSL first constructs the distributed storage of data graph $d$, and then lists all matches of pattern graph $p$ in data graph $d$ from scratch.
\item \textbf{Incremental Updating.} In this stage, according to the change of $d$, DDSL first updates the distributed storage of $d$, and then updates the matches of $p$.
\end{enumerate}

\begin{figure}[htbp]
\centering
\includegraphics[width=\linewidth, keepaspectratio]{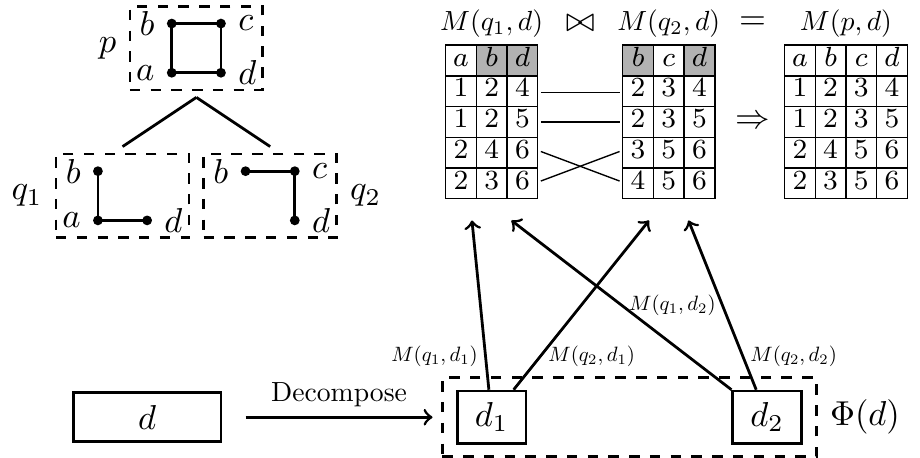}
\caption{Distributed join framework.}
\label{fig:framework}
\end{figure}

The initial calculation follows a general distributed join framework \cite{lai2016scalable}, which is shown in Fig. \ref{fig:framework}. It has two main building blocks: a distributed storage mechanism, and a set of join units. Firstly, according to the distributed storage mechanism, the data graph $d$ is decomposed into several parts ($d_1$, $d_2$). Then the pattern graph $p$ is decomposed into several join units ($q_1$, $q_2$), and the matches of each unit can be directly listed from each part of $d$ without join. By gradually joining the matches of all units together, we can obtain the matches of the pattern graph.

Note that, all partial matches generated in this framework are first saved to disks and then loaded for join. Thus, those matches are treated as intermediate results, and are counted into the total cost. In Fig. \ref{fig:framework}, matches of $q_1$ and $q_2$ ($M(q_1, d)$ and $M(q_2, d)$, resp.) are intermediate results. Actually, if we view the join order as a tree structure (called the join tree), all nodes except for the root correspond to a match set which belongs to the intermediate results.

Different storage mechanisms may support the direct listing of matches of different join units. For example, two different storage mechanisms have been proposed in \cite{lai2015scalable,lai2016scalable} to support stars and/or cliques as join units. Intuitively, as Example \ref{example:jointrees} shows, supporting more general units offer more flexibility in finding lower join cost solution. In DDSL we use a distributed storage mechanism called \textit{Neighbor-Preserved} (NP) storage, which supports all graphs with radius=$1$ as join units (called \textbf{R1 units}). Moreover, this storage mechanism has a bounded space cost.

\begin{figure}[htbp]
    \centering
    \subfloat[triangles]{
        \includegraphics[width=.25\linewidth, keepaspectratio]{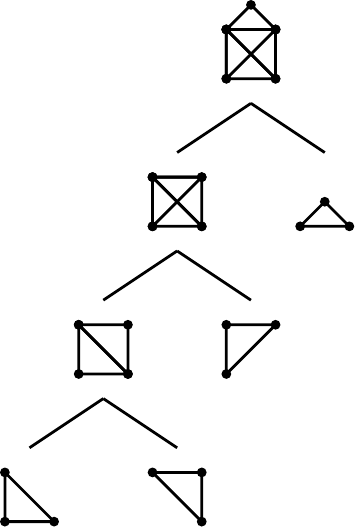}
        \label{fig:jointrees_a}
    }%
    \subfloat[cliques]{
        \includegraphics[width=.25\linewidth, keepaspectratio]{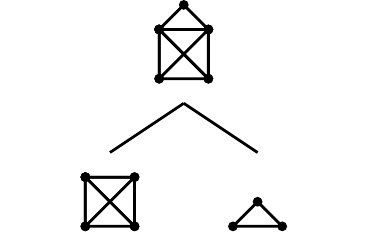}
        \label{fig:jointrees_b}
    }%
    \subfloat[R1 units (no join)]{
        \includegraphics[width=.25\linewidth, keepaspectratio]{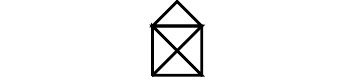}
        \label{fig:jointrees_c}
    }%
    \caption{Join trees of a typical pattern using different join units.}
    \label{fig:jointrees}
\end{figure}

\begin{example}
\label{example:jointrees}
Fig. \ref{fig:jointrees} shows three different join trees of a typical pattern. In Fig. \ref{fig:jointrees_a}, each join unit is a triangle, so the pattern graph is decomposed into $4$ triangles. In Fig. \ref{fig:jointrees_b} and Fig. \ref{fig:jointrees_c}, the join units are cliques and R1 units, respectively. Note that a triangle is a special clique of size $3$, and a clique is a special R1 unit. In this example, using more general join units can reduce the number of joins. Specifically, using cliques can reduce the number of joins by $2$ compared with using triangles, and using R1 units requires no join at all. The intermediate result size is also reducing in this case. For example, in Fig. \ref{fig:jointrees_b}, the intermediate pattern set is a subset of that in Fig. \ref{fig:jointrees_a}, so the intermediate result size in Fig. \ref{fig:jointrees_b} is less than that in Fig. \ref{fig:jointrees_a}. In Fig. \ref{fig:jointrees_c}, the intermediate result size is $0$.
\end{example}

To further reduce the I/O and communication cost of the join operations, we incorporate the idea of match compression into our method. Specifically, we use the vertex-cover-based compression (VCBC) \cite{qiao2017subgraph} to compress all the matches of join units. 
A nice property of the VCBC is that, we can perform join operations directly on the compressed data without decompression, which saves the computation cost.
Besides the compression, we derive a comprehensive cost model to estimate the join cost of any given join tree. Based on this cost model, we use a dynamic programming to find the optimal join tree which has the minimum join cost.

In the incremental updating stage, we design an algorithm which can update the NP storage according to the graph changes with a low cost. To update the matches, our idea is to extract a patch set that contains only the newly-appeared matches. Then we can simply merge the old matches with the patch set, and filter out the matches that no longer exist. However, computing the patch set using the regular join is no easier than computing all matches from scratch, because we may not effectively filter out the unnecessary matches on either join side. In DDSL, we design a novel \textit{Navigated Join} (Nav-join) to extract the patch set with a lower cost. We first decompose the pattern graph into a left-deep tree, and then compute a partial patch set using the Nav-join along this tree. For each involved join, we partition the matches on the left side to the corresponding machines and perform a local exploration, which saves the I/O cost for listing and loading the matches on the right side. The final patch set is then obtained by merging several partial patch sets computed through a set of carefully-picked trees. 

In summary, we make the following contributions.
\begin{itemize}
    \item To the best of our knowledge, DDSL is the first approach that supports unlabeled and undirected pattern matching on dynamic graphs in a distributed environment.
    \item We derive a comprehensive cost model for estimating the cost of a given join tree, and then design a dynamic programming algorithm to find the optimal join tree which has the lowest estimating cost.
    \item To handle dynamic graphs, we design an efficient algorithm update the NP storage within a single MapReduce round. We then propose the novel Nav-join to efficiently compute the patch set. By merging the patch set with the old match set, and filtering out invalid matches, we can obtain the updated result.
    \item We conduct extensive experiments on real-world graphs to demonstrate the efficiency of DDSL on both static and dynamic graphs.
\end{itemize}

\subsection{Paper Organization}
The rest of this paper is organized as follows. In Section \ref{sec:preliminaries}, we introduce the basic concepts in this paper. In Section \ref{sec:initialcalc}, we describe the underlying distributed storage and how we do the initial calculation in DDSL. Then in Section \ref{section:matchcompression} and Section \ref{section:optimaltree}, we propose several optimizations to reduce the cost of initial calculation. To handle dynamic graphs, we propose the incremental updating algorithm in Section \ref{section:incremental}. Section \ref{sec:related} surveys the related works and compares them to our work. In Section \ref{section:experiments}, we conduct extensive experiments on real-world networks, to show the efficiency of our approach on both static and dynamic graphs. Finally, we conclude in Section \ref{sec:conclusion}.

\section{Preliminaries}
\label{sec:preliminaries}
In this section we introduce the basic concepts and definitions. We also list important notations used in this paper as well as their meanings in Table \ref{table:notations}.
\begin{small}
\begin{table}
\centering
\caption{Notations}
\label{table:notations}
\begin{tabular}{|c|c|}
\hline
Symbol         & Description                           \\
\hline
$d$, $p$       & The data graph $d$ and pattern graph $p$. \\
\hline
$v_i$, $u_i$   & A vertex which has id $i$. \\
\hline
$(v_i, v_j)$   & An edge that connects $v_i$ and $v_j$. \\
\hline
$(v_i, v_j, v_k)$ & A triangle formed by $v_i$, $v_j$ and $v_k$. \\
\hline
$V(g)$, $E(g)$ & The set of vertices and edges in $g$. \\
\hline
$\neighbor[g]{v}$ & The neighbor set of $v$ in $g$.       \\
\hline
$deg(v)$       & The degree of $v$.                      \\
\hline
$g[V']$        & The subgraph of $g$ induced by $V'$.    \\

\hline
$f$            & A match of $p$ in $d$.                  \\
\hline
$q$            & A join unit.                            \\
\hline
$M(p,d)$       & The match set of $p$ on $d$.             \\
\hline
$M_{ac}(q, d_i)$ & The anchor-center-constrained match set of $q$. \\
\hline
$\mathcal{T}(p)$ & The optimal join tree of $p$.          \\
\hline
$V_c(p)$       &  A vertex cover of $p$.                  \\
\hline
$s$            & A skeleton(a partial match from $V_c(p)$ to $V(d)$).  \\
$f|s$          & The compressed form of all matches having $s$. \\

\hline
$loc(u)$       & The local graph of vertex $u$.           \\
\hline
$d_i$          & A part of $d$ in the distributed storage. \\
\hline
$\Phi(d)$      & The Neighbor-Preserved storage of $d$.    \\

\hline
$E_a(U)$,$E_d(U)$ & The set of edges to be added/deleted. \\
\hline
$M_{new}(p, d')$ & The patch set $M(p, d')\backslash M(p, d)$. \\
\hline
\end{tabular}
\end{table}
\end{small}

\subsection{Graph and Subgraph}
Given a graph $g$, we use $V(g)$ and $E(g)$ to denote the vertex set and edge set in $g$, respectively. Each vertex $v \in V(g)$ is associated with a unique id $i$, and thus is denoted by $v_i$. An edge connecting $v_i$ and $v_j$ is denoted by $(v_i, v_j)$. Edge $(v_i, v_j)$ is \textit{incident} to both $v_i$ and $v_j$. In this paper we focus on undirected and unlabeled graphs, which means $(v_i, v_j)$ is identical to $(v_j, v_i)$.

For any vertex $v \in V(g)$, denote $\neighbor{v}$ as the \textit{neighbors} of $v$, which is defined as $\neighbor{v}=\{u|(v, u) \in E(g)\}$, and $deg(v)=|\neighbor{v}|$ as the degree of $v$.

Given two graphs $g_1$ and $g_2$, $g_2$ is a \textit{subgraph} of $g_1$ if $V(g_2) \subseteq V(g_1)$ and $E(g_2) \subseteq E(g_1)$. Specifically, $g_2$ is the subgraph of $g_1$ \textit{induced by} vertex set $V'$ if (1) $V(g_2)=V' \subseteq V(g_1)$, and (2) $E(g_2)=E(g_1) \cap (V' \times V')$. In this case we also denote $g_2$ by $g_1[V']$. Apparently any graph $g$ is the subgraph of itself induced by $V(g)$.

\subsection{Subgraph Listing}
By introducing the concept of graph isomorphism, we can then formally define the subgraph listing problem.

\begin{definition}[Graph Isomorphism \cite{west2001introduction}]
Given two graphs $g_1$ and $g_2$, an isomorphism from $g_1$ to $g_2$ is a bijection $f: V(g_1) \mapsto V(g_2)$ such that $(v_i, v_j) \in E(g_1)$ if and only if $(f(v_i), f(v_j)) \in E(g_2)$. If there is an isomorphism from $g_1$ to $g_2$, then we say $g_1$ is isomorphic to $g_2$.
\end{definition}

\begin{definition}[Subgraph Listing]
Given two connected graphs $g_1$ and $g_2$, subgraph listing requires outputing all subgraphs of $g_1$, which are isomorphic to $g_2$. Here $g_1$ is also called the \textit{data graph}, and $g_2$ is also called the \textit{pattern graph}.
\end{definition}

For the ease of description, in the rest of this paper, we use $d$ to denote the data graph, and $p$ to denote the pattern graph. Each isomorphism $f$ from $p$ to a valid subgraph of $d$ is called a \textit{match}. Next we introduce the automorphism of a graph. As it introduces duplicate matches and harms the efficiency of the algorithm, we also describe the technique to break it.

The \textit{automorphism} of a graph $g$ is the isomorphism $f: V(g) \mapsto V(g)$. Any graph has an automorphism $f(v_i)=v_i$, while some graphs have more. In subgraph listing, if the pattern graph $p$ has $k$ automorphisms, one will find $k$ matches $p$ to each valid subgraph of $d$. Extra effort has to be paid to remove such duplicates, and thus will harm the efficiency. A common way to avoid duplicate results is \textit{Symmetric Breaking} (SimB) \cite{grochow2007network}. It assigns a partial order $\mathbf{ord}=\{v_i \prec v_j\}$ for some node pairs $\pair{v_i}{v_j} \in V(p) \times V(p)$, and a total order $\{u_i < u_j \text{ if } i < j\}$ for all node pairs $\pair{u_i}{u_j} \in V(d) \times V(d)$. Then an isomorphism $f$ is valid if and only if $f(v_i) < f(v_j)$ for any $v_i \prec v_j$. By carefully assigning the partial order, SimB guarantees that for any graph $g$ which is isomorphic to $p$, there is only one valid isomorphism from $p$ to $g$. In the following of this paper, we assume that SimB has been applied, and for each valid subgraph of $d$, we can find only one match. In this case, listing all valid subgraphs is equal to listing all valid matches, which is denoted as $M(p,d)$.

\subsection{Graph Update}
For a general undirected and unlabeled graph, there are $4$ types of updates:
\begin{itemize}
    \item Adding a vertex;
    \item Deleting an existing vertex, as well as all edges incident to it;
    \item Adding an edge between two existing vertices;
    \item Deleting an existing edge.
\end{itemize}
Since we require $d$ and $p$ to be connected, we only consider edge insertion/deletion in this paper. Adding a vertex $u$ can be automatically done when we add an edge $(u, v)$, and deleting $u$ can be simulated by first deleting all edges incident to $u$, then deleting $u$ automatically.

One way to handle graph changes is to consider the changes as a stream, and to only deal with one single insertion/deletion at a time. Another way is to consider the changes as several batches, where each batch contains several edges to be inserted and deleted simultaneously. In this paper we use batch updates for two reasons. First, by considering a stream as batches with only one edge, the solution on batch updates can also apply on stream. Second, each time we update the graph and the result, some basic cost like reading the whole graph is unavoidable, and by dealing with multiple edges simultaneously we may save such cost. Under this setting, a \textit{graph update} $U$ contains two edge sets $E_d(U)$ and $E_a(U)$, where $E_d(U)$ contains edges to be deleted and $E_a(U)$ contains edges to be added.

\subsection{Dynamic Graph and Problem Statement}
A dynamic graph is defined as an initial graph $d$, followed by several updates $U_1$, $U_2$, $\ldots$. Assuming that after applying each of those updates, we get the updated graphs $d'$, $d''$, $\ldots$, our goal is to output the match set $M(p, d)$, $M(p, d')$, $M(p, d'')$, $\ldots$ efficiently. Here we take a simplified version: only one update exists. The reason is that, if we can compute $M(p, d')$ efficiently, we can handle the next one by treating $d'$ as the new initial graph. We then describe the target problem as follows.

\textbf{Problem Statement.} Given a distributed file system, which can store and load data as key-value pairs, a pattern graph $p$, a data graph $d$, and a graph update $U$, the problem has two goals:
\begin{enumerate}
    \item To compute $M(p, d)$.
    \item To compute $M(p, d')$ given $M(p, d)$, where $d'$ is the new graph after applying $U$ on $d$.
\end{enumerate}

\subsection{Power-Law Random Graph Model}
In this paper, we use the \textit{Power-Law Random} (PR) graph model \cite{chung2003eigenvalues} to analyze the data graph $d$. Based on it, we build a comprehensive estimator to estimate the number of matches given any pattern graph $p$ and data graph $d$.

It is known that in most real-life networks, the degree of vertices follows the power-law distribution. Thus, modeling the underlying graph as a PR graph can usually give a realistic estimation of its properties. In a PR graph $d$, edge $(v_i, v_j)$ is assigned with probability $\Pr(i,j)=deg(v_i) \cdot deg(v_j) \cdot \rho$, where $\rho=\frac{1}{2|E(d)|}$. We can verify that $\mathbb{E}(deg(v_i))=deg(v_i)$ if two assumptions hold. One is that self-loop is allowed, and another is that for any $v_i$, $deg(v_i) \leq \sqrt{2|E(d)|}$. Although not all graphs satisfy these two conditions, this model still serves as a good guidance in practice \cite{lai2015scalable}.

\section{Initial Calculation}
\label{sec:initialcalc}


In the initial calculation, the task is to perform regular subgraph listing on the data graph $d$. As discussed, in this stage we follow a distributed join framework, which has two basic building blocks:
\begin{enumerate}
\item The set of supported join units.
\item The distributed storage mechanism of $d$ to support the direct listing of matches of each join unit.
\end{enumerate}
In this section we describe the details of these two building blocks that are used in DDSL.

\subsection{R1 Units}
\label{section:unitandstorage}
As Example \ref{example:bushycost} shows, using more general join units may reduce the intermediate result size. Thus, instead of using only one or two kinds of join units, in DDSL we use all graphs with \textit{radius=1} as our join units, which are called R1 units. They have the following advantages:
\begin{itemize}
    \item R1 units contain a variety of shapes, and thus has the potential to reduce the join cost.
    \item Using R1 units allows us to compress the join results (see Section \ref{section:matchcompression}), and most importantly, to incrementally update the join results (see Section \ref{section:incremental}).
\end{itemize}

\begin{definition}[R1 Units]
A graph $q$ is an R1 unit if and only if there exists a vertex $v$ in $V(q)$ s.t. $\neighbor[q]{v} \cup \{v\} = V(q)$.
\end{definition}

\textbf{Anchor Vertex.} Given an R1 unit $q$, we can find at least one vertex, who is the common neighbor of all other vertices. We randomly pick such a vertex as the \textit{anchor vertex} of $q$. Later we will utilize the anchor vertex to facilitate the listing and join process.

\subsection{Neighbor-Preserved Storage}
In the join-based framework, matches of each join unit are listed from each part of the graph storage \textit{completely and independently}. Suppose $d$ is stored in $m$ parts $d_1, d_2, \ldots, d_m$, and $q$ is a join unit, then the completeness requires
\begin{small}
\begin{align}
    M(q,d)\subseteq \bigcup_{i \in [1,m]}M(q, d_i),
\end{align}
\end{small}
that is, every match exists in at least one partition and no match will be lost. The independence requires
\begin{align}
    \forall_{1\leq i < j \leq m}: M(q, d_i) \cap M(q, d_j)= \emptyset,
\end{align}
that is, no duplicate matches are found in different partitions. If these two properties are satisfied, we can obtain $M(q, d)$ by simply adding all matching set $M(q, d_i)$ together.

To support directly listing the matches of any R1 unit, we use the Neighbor-Preserved (NP) storage mechanism. An NP storage $\Phi(d)$ is built upon an arbitrary partition function $h$, and the \textit{local graph} $loc(u)$ of each vertex $u \in V(d)$.

\begin{definition}[Partition Function]
A partition function $h$ maps vertex ids to partition ids. Given any vertex id, $h$ outputs a partition id $j \in [1, m]$, where $m$ is the number of partitions.
\end{definition}

\begin{definition}[Local Graph]
A local graph $loc(u)$ of vertex $u \in V(d)$ is the induced graph of vertex set $\{u\} \cup \neighbor[d]{u}$, i.e., $loc(u)=d[\{u\} \cup \neighbor[d]{u}]$.
\end{definition}

Suppose $h(i)=j$ for a vertex $u_i \in V(d)$, the NP storage stores $loc(u_i)$ in $d_j$. $d_j$ is thus defined as
\begin{small}
\begin{align*}
d_j=\bigcup\limits_{\forall u_i \in V(d), h(i)=j} loc(u_i).
\end{align*}
\end{small}
Perceptually $u_i$ lies on the ``center'' of $loc(u_i)$, so we call $u_i$ a \textit{center vertex} of $d_j$ if $h(i)=j$. For non-center vertices of $d_j$, we call them \textit{border vertices}.

One can easily verify that $M(q, d_i)$ does not satisfy the independence requirement, however, we can use an alternative version $M_{ac}(q, d_i)$ which satisfies the anchor-center constraint.
We first describe how $M_{ac}(q, d_i)$ is computed, and then proof the independence and completeness of it in Lemma \ref{lemma:joinunit}. Given $q$ and $d_i$, $M_{ac}(q, d_i)$ is computed by listing all matches from $q$ to $d_i$, under the constraint that $q$'s anchor vertex must be matched to one of $d_i$'s center vertices. This can be done using existing in-memory subgraph listing algorithms with minor modifications. For example, in a DFS algorithm, we can put $q$'s anchor vertex in the first level, and only try to match it to $d_i$'s center vertices.





\begin{lemma}
\label{lemma:joinunit}
With $M_{ac}(q, d_j)$, any R1 unit can be the join unit w.r.t. an NP storage.
\begin{proof}
We proof this by showing the independence and completeness of $M_{ac}(q, d_i)$.

\textbf{Independence.} We proof the independence by showing that any two different sets have no common matches. Let $q$'s anchor vertex be $v_0$, and $f$, $f'$ be two arbitrary matches in $M_{ac}(q, d_i)$ and $M_{ac}(q, d_j)$, respectively. Suppose $f(v_0)=u_x$ and $f'(v_0)=u_y$, which means $h(x)=i$ and $h(y)=j$. Since $h$ is a partition function, $i \ne j \implies x \ne y$, and thus $f \ne f'$.

\textbf{Completeness.} We further prove the completeness by showing that any valid match in $M(q, d)$ must exist in one $M_{ac}(q, d_i)$. Let $q$'s anchor vertex be $v_0$, and $f(v_0)=u_x$. By the definition of an anchor vertex, we have
\begin{align*}
\forall v_j \in V(q): v_j \in \{v_0\} \cup \neighbor[q]{v_0},
\end{align*}
and thus
\begin{align*}
\forall (v_j, v_k) \in E(q): (f(v_j), f(v_k)) \in E(loc(u_x)) \subseteq E(d_{h(x)}).
\end{align*}

We proved that $f \in M(q, d_{h(x)})$, and we also know $u_x$ is a center vertex in $d_{h(x)}$. Combining these two we have $f \in M_{ac}(q, d_{h(x)})$.
\end{proof}
\end{lemma}

To build the NP storage $\Phi(d)$, we store some duplicate edges between border vertices in each part. Then the important question is: whether the space cost of $\Phi(d)$ can be bounded? We analyze and give a confirmative answer to this question w.r.t. the extra edges needed to build $\Phi(d)$.

We start with an arbitrary edge $(u_x, u_y)$. In $\Phi(d)$, $(u_x, u_y)$ is stored in $loc(u_x)$, $loc(u_y)$ and $loc(u_z), \forall u_z \in \neighbor[g]{u_x} \cap \neighbor[g]{u_y}$, where such edges stored in each $loc(u_z)$ are the extra edges introduced to close the triangle $(u_x, u_y, u_z)$. In other words, for each triangle $(u_x, u_y, u_z) \in d$, three extra edges $(u_x, u_y)$, $(u_x, u_z)$ and $(u_y, u_z)$ are stored in $loc(u_z)$, $loc(u_y)$ and $loc(u_x)$, respectively. When we take the union of local graphs that belong to the same $d_i$, some edges might be merged. For example, if $loc(u_x)$ and $loc(u_y)$ are both stored in $d_j$, then $(u_x, u_z)$ and $(u_x, u_y)$ are each stored only once, saving two edges. In summary, the total extra edges introduced to build $\Phi(d)$ is bounded by $3\cdot\Delta(d)$, where $\Delta(d)$ is the number of triangles in $d$.

The $3\cdot\Delta(d)$ bound is suitable for a sparse $d$ or a large $m$. When $d$ is sparse, $\Delta(d)$ is small. When $m$ is large, for each triangle in $d$, there are few chances that two or three of its vertices are center vertices of the same $d_j$. In both cases there are only a few edges that are merged, so $3\cdot\Delta(d)$ is close to the true extra space cost of $\Phi(d)$. However, when $d$ is dense, or $m$ is very small, there are many merged edges, and the gap between $3\cdot\Delta(d)$ and the true extra cost is large. In the most extreme case, when $d$ itself is a clique, and $m=1$, the true extra cost is $0$, while $\Delta(d) \approx |V(d)|^3$. We thus bound the extra cost in another way. Since $d_i$ cannot be larger than $d$, we simply bound the extra cost as $(m-1)\cdot E(d)$. Noticed that both bounds are hold for all situations, thus in summary, the bound is $\min\{3\cdot\Delta(d), (m-1)\cdot E(d)\}$.

\subsection{Pattern Decomposition and Join}
With the two building blocks decided, we can perform the subgraph listing within the join framework. Given a pattern graph $p$, we decompose $p$ into a set of join units $Q=\{q_1, q_2,\ldots,q_k\}$, such that:
\begin{enumerate}
\item Each $q_i$ is an R1 unit;
\item $\bigcup\limits_{q_i \in Q} q_i = p$.
\end{enumerate}
This decomposition can be done by simply generating all possible R1 units inside $p$, and then exhaustively search a subset that satisfies the second condition.

After listing the matches of all the join units in $Q$, we then need to join the matches together. Assuming that we have two matches $f_1 \in M(p_1, d)$ and $f_2 \in M(p_2, d)$, then $f_1$ can be joined with $f_2$ if $\forall v \in V(p_1)\cap V(p_2): f_1(v)=f_2(v)$. This guarantees that there exists no conflict when merging the mappings in $f_1$ and $f_2$, and the vertex set $\{f_i(v)|\forall v \in V(p_1)\cap V(p_2)\}$ is called the join keys. In Fig. \ref{fig:framework}, we illustrate a simple join between $M(q_1, d)$ and $M(q_2, d)$, where the columns to generate join keys are marked in gray. After the join, we need to check each result, and drop every match that (1) maps two or more vertices in $p$ to the same vertex in $d$, or (2) violates the partial order $\mathbf{ord}$.

Next we will discuss several optimizations to reduce the cost in this framework.

\section{Match Compression}
\label{section:matchcompression}
In general, given any order $\langle v_1, v_2, \ldots, v_r \rangle$ of $V(p)$, any match $f$ can be stored in a plain form $\langle f(v_1), f(v_2), \ldots, f(v_r) \rangle$.  However, such a plain storage may occupy a lot of space. For example, the authors in \cite{qiao2017subgraph} find that for a $6$-vertex pattern and a data graph with $0.1$ million vertices, the match set can take $10^4$ petabytes of storage. They also show that applying match compression can significantly reduce the storage size as well as the I/O cost. Thus, it is worth incorporating the match compression technique into our method.

The idea of match compression is to use a compressed form to store multiple matches together, in order to save the cost of disk I/O and network communication cost. There are different compression strategies in the literature, and in DDSL we use the vertex-cover-based compression (VCBC) \cite{qiao2017subgraph}, because it can be incorporated into the join process without introducing extra I/O cost of decompression. To make the paper self-contained, we first explain the compression and decompression of VCBC in Section \ref{section:compression} and Section \ref{section:decompression}, then discuss its compression ratio in Section \ref{section:optimalcompression}. In our analysis, we assume that each vertex can be stored in an integer, and use the number of integers to measure the storage size. For example, if we store all matches in the plain form, then for $k$ matches we need to store $k\cdot r$ integers in total.

\subsection{Compression}
\label{section:compression}
Let $V_c(p)$ be a vertex cover of $p$, and $s:V_c(p) \mapsto V(d)$ be a partial match w.r.t. $V_c(p)$, which is called a \textit{skeleton}. For any match $f_i$, it's skeleton $s_i$ w.r.t. $V_c(p)$ can be easily computed as $\{u \mapsto f_i(u)|\forall u \in V_c(p)\}$. The vertex-cover-based compression stores all matches who have the same skeleton $s$, namely $M|s=\{f_i|s_i=s\}$, in a compressed form $f|s$. For any $v_x \in V_c(p)$, all matches in $M|s$ maps $v_x$ to the same vertex in $d$, so $f|s$ only maps $v_x$ to one vertex. For other $v_y \in V(p)\backslash V_c(p)$, matches in $M|s$ may map it to many different vertices, so $f|s$ maps $v_y$ to a vertex set $\{f_i(v_y)|s_i=s\}$. The vertices in $V(p)\backslash V_c(p)$ are thus referred to as the \textit{compressed vertices} w.r.t. $V_c(p)$. Given a typical pattern $p$, the skeleton $c$, and the conditional match set $M|s$, Lemma \ref{lemma:compressionratio} shows that, the storage saved by $f|s$ has a lower bound.

\begin{figure}[htbp]
    \centering
    \subfloat[no compression]{
        \includegraphics[width=.32\linewidth, keepaspectratio]{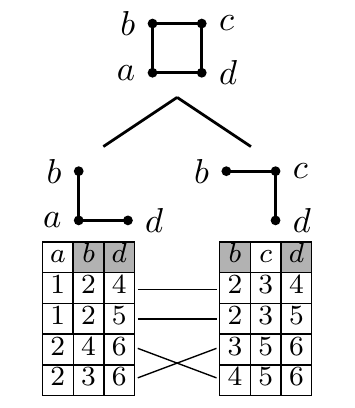}
        \label{fig:join_a}
    }%
    \subfloat[VCBC with $\{a, b, c\}$]{
        \includegraphics[width=.32\linewidth, keepaspectratio]{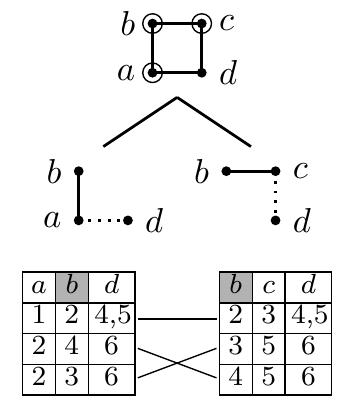}
        \label{fig:join_b}
    }%
    \subfloat[VCBC with $\{a, c\}$]{
        \includegraphics[width=.32\linewidth, keepaspectratio]{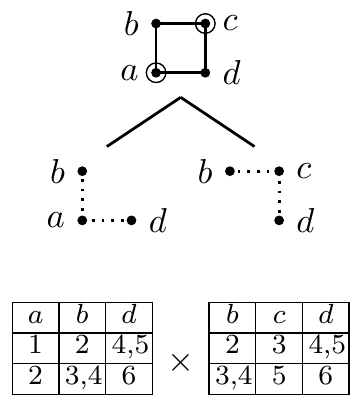}
        \label{fig:join_c}
    }%
    \caption{Matches and joins under different compression strategies.}
    \label{fig:join}
\end{figure}

\begin{example}
\label{example:compression}
Fig. \ref{fig:join} compares the matches and joins under three compression strategies. Each row in the table is a match of the above pattern, and the columns of join keys in each table are marked in gray. In Fig. \ref{fig:join_a}, no compression is applied. In Fig. \ref{fig:join_b}, vertex $d$ is the compressed vertex, so match $\langle 1, 2, 4 \rangle$ and $\langle 1, 2, 5 \rangle$ are compressed into $\langle 1, 2, \{4, 5\} \rangle$. Similarly match $\langle 2, 3, 4 \rangle$ and $\langle 2, 3, 5 \rangle$ are compressed into $\langle 2, 3, \{4, 5\} \rangle$. In Fig. \ref{fig:join_c}, another vertex $b$ becomes a compressed vertex, so more matches are compressed.
\end{example}

\begin{lemma}
\label{lemma:compressionratio}
Suppose there are $a$ vertices in $V_c(p)$, $b$ vertices in $V(p)\backslash V_c(p)$, and $c$ matches in $M|s$, then $f|s$ saves at least $a\cdot(c-1)$ integers in storage.
\begin{proof}
The uncompressed form stores $c\cdot(a+b)$ integers. For each match, $f|s$ needs to store at most $b$ integers, one for each vertex in $V(p)\backslash V_c(p)$. Thus, in total $f|s$ needs to store less than $a+b\cdot c$ integers, which saves at least $a\cdot(c-1)$ integers compared to the uncompressed form.
\end{proof}
\end{lemma}



\subsection{Decompression}
\label{section:decompression}
The decompression of a compressed match $f|s$ can be done in two steps:
\begin{enumerate}[i.]
\item For all $v \in V(p)\backslash V_c(p)$, take the Cartesian product of the vertex sets they are mapped to. Each tuple plus the skeleton $s$ becomes a candidate match;
\item For each candidate match, drop it if: (1) it is not an injection, i.e., maps two vertices in $p$ to the same one in $d$; or (2) it violates the partial order $\mathbf{ord}$.
\end{enumerate}
After these two steps, the remaining matches are the recovered matches.

\begin{definition}[Correct Compressed Match]
Given a compressed match $f|s$ of $p$, $f|s$ is \textit{correct} if the following condition holds
\begin{enumerate}
    \item $\forall f \in M(p, d)$, which has skeleton $s$, $f$ can be recovered from $f|s$;
    \item $\forall f$ which is decompressed from $f|s$, $f \in M(p, d)$.
\end{enumerate}
\end{definition}

It is proved that the vertex-cover-based compression in Section \ref{section:compression} is correct \cite{qiao2017subgraph}.

\subsection{The Optimal Compression}
\label{section:optimalcompression}
A pattern graph can have numerous vertex covers, where each corresponds to a unique compressed match set. Lemma \ref{lemma:compressionratio} gives the lower bound of a specific match set $M|s$, however, usually we care more about the total compression ratio $R$ given the pattern $p$ and the vertex cover $V_c(p)$. Computing the precise ratio requires to compute all matches, which are unavailable unless we finish the subgraph listing task. Therefore, we try to compute a lower bound of $R$ with less information. Theorem \ref{theorem:lowestcompression} gives a lower bound of $R$, given the size of $M(p, d)$ and $M(p[V_c(p)], d)$.

\begin{theorem}
\label{theorem:lowestcompression}
Given $p$, $V_c(p)$, and $d$, the lower bound of the compression ratio is
\begin{small}
\begin{align}
&R_{lower} = \nonumber \\
&\frac{|V(p)|\cdot |M(p,d)|}{|V(p)|\cdot |M(p,d)| + |V_c(p)|\cdot (|M(p[V_c(p)],d)| - |M(p,d)|)}.\label{equation:ratio}
\end{align}
\end{small}
\begin{proof}
The uncompressed form stores each match separately, so the storage needed is
\begin{align}
S_{plain}=|V(p)|\cdot |M(p,d)|. \label{equation:plainstorage}
\end{align}

We break the compressed storage into two parts: the skeleton part and the compressed part.

Each skeleton is a match of the induced graph $p[V_c(p)]$ in $d$, so the number of distinct skeletons is no greater than $|M(p[V_c(p)], d)|$. As skeletons are stored separately in each $f|s$, so the maximum storage required is:
\begin{align}
S^{max}_{skeleton} = |V_c(p)|\cdot |M(p[V_c(p)],d)|. \label{equation:skeletonstorage}
\end{align}
Following Lemma \ref{lemma:compressionratio}, the compressed form stores at most $|V(p)| - |V_c(p)|$ integers for each match, so the maximum number of integers stored by the compressed part is
\begin{align}
S^{max}_{compress}=(|V(p)| - |V_c(p)|)\cdot |M(p, d)|. \label{equation:compressstorage}
\end{align}
Combining Equation \ref{equation:plainstorage}, Equation \ref{equation:skeletonstorage} and Equation \ref{equation:compressstorage}, we have the lower bound of the compression ratio:
\begin{small}
\begin{align*}
R & =\frac{S_{plain}}{S_{skeleton} + S_{compress}}  \\
& \geq \frac{S_{plain}}{S^{max}_{skeleton} + S^{max}_{compress}}  \\
& = \frac{|V(p)|\cdot |M(p,d)|}{|V(p)|\cdot |M(p,d)| + |V_c(p)|\cdot (|M(p[V_c(p)],d)| - |M(p,d)|)}. 
\end{align*}
\end{small}
\end{proof}
\end{theorem}

Currently the actual values of $|M(p,d)|$ and $|M(p[V_c(p)],d)|$ are unknown, and computing these two values is NP-hard. In this case we introduce an estimator of these two values in Section \ref{section:estimation}. For any pattern $p$, there exists a vertex cover which gives the highest $R_{lower}$, Such a compression provides the best guaranteed compression ratio, and thus is called the \textit{optimal compression}.

\subsection{Match Size Estimation}
\label{section:estimation}
In order to estimate $R_{lower}$ w.r.t. a given vertex cover $V_c(p)$ of $p$, we need to estimate the total number of matches of $p$ in $d$. To do so, we model the data graph $d$ as a PR graph, and compute the expected match size, which is denoted by $\mathrm{E}(|M(p, d)|)$.

Our computation is based on a random-assignment process. In this process, we first randomly assign a match $f$ from $V(p)$ to $V(d)$. Then we test whether $f$ is a valid match. If $f$ is a valid match, then it satisfy two conditions:
\begin{enumerate}[MC$_1$]
    \item For any mapping $v \mapsto u$ in $f$, $deg(v) \leq deg(u)$, where $deg(v)$ is the degree of $v$ in $p$, and $deg(u)$ is the degree of $u$ in $d$;
    \item For any edge $(v_i, v_j) \in E(p)$, there exists an edge $(f(v_i), f(v_j))$ in $E(d)$.
\end{enumerate}

We first consider a simplified version of MC$_1$, which is denoted as MC$_0$. Given $V(p)=\{v_1, v_2, \ldots, v_k\}$, and a degree sequence $W=\langle w_1, w_2, \ldots, w_k \rangle$, where $\forall w_i: w_i \geq deg(v_i)$, condition MC$_0$ restricts $\forall i: deg(f(v_i))=w_i$, i.e., the degree of vertex $f(v_i)$ in $d$ must be $w_i$. Let $\mathrm{X_0}$, $\mathrm{X_1}$ and $\mathrm{X_2}$ be the event that MC$_0$, MC$_1$ and MC$_2$ are satisfied, respectively, then we have
\begin{small}
\begin{align*}
\mathrm{Pr}(X_0|f, W)=\prod_{v_i \in V(p)} \mathrm{p}_{w_i},
\end{align*}
\end{small}
and
\begin{small}
\begin{align*}
\mathrm{Pr}(X_2|X_0)=\prod_{(v_i, v_j) \in E(p)}w_i w_j \rho.
\end{align*}
\end{small}
Note that in the formula of $\mathrm{Pr}(X_2|X_0)$, term $w_i$ appears once for each adjacent edge of $v_i$, so in total $w_i$ appears $deg(v_i)$ times. Therefore, we can write
\begin{small}
\begin{align}
\mathrm{Pr}(X_2|X_0)=\rho^{E(p)}\cdot\prod_{v_i \in V(p)}w_i^{deg(v_i)}.
\end{align}
\end{small}
Thus, given $W$, the probability that $f$ is valid is
\begin{small}
\begin{align*}
\mathrm{Pr}(X_0, X_2|f, W)&=\mathrm{Pr}(X_2|X_0)\cdot\mathrm{Pr}(X_0|f, W) \\
&=\rho^{E(p)}\cdot\prod_{v_i \in V(p)}w_i^{deg(v_i)}\mathrm{p}_{w_i}.
\end{align*}
\end{small}
Recall that MC$_0$ is a special case of MC$_1$, so $f$ satisfies MC$_1$ if there exists any one $W$ s.t. $f$ satisfies MC$_0$. Therefore, the probability that $f$ satisfies MC$_1$ and MC$_2$ is
\begin{small}
\begin{align}
\epsilon&=\mathrm{Pr}(X_1, X_2|f) \nonumber \\
&=\sum_{W}\mathrm{Pr}(X_0, X_2|f, W) \nonumber\\
&= \sum_{w_1=deg(v_1)}^{\infty}\sum_{w_2=deg(v_2)}^{\infty}\cdots\sum_{w_k=deg(v_k)}^{\infty}\mathrm{Pr}(X_0, X_2|f, W) \nonumber\\
&=\rho^{|E(p)|}\cdot \prod_{v_i \in V(p)}\sum_{w=deg(v_i)}^{\infty}w^{deg(v_i)}\mathrm{p}_w \label{equation:validmatchprobability}
\end{align}
\end{small}
Assuming that the pattern graph $p$ has $O(1)$ vertices, the term $\sum_{i=deg(v)}^{\infty}i^{deg(v)}\mathrm{p}_i$ only has $O(1)$ possible values, thus we can compute those values in advance, and then Equation \ref{equation:validmatchprobability} can be evaluated in $O(1)$ time.

Then $\mathrm{E}(|M(p, d)|)$ is calculated as $\epsilon$ times the number of all possible assignments, then divided by the number of automorphisms of $p$ satisfying $\mathbf{ord}$, which is
\begin{small}
\begin{align}
\mathrm{E}(|M(p, d)|)=\frac{|V(d)|!}{(|V(d)|-|V(p)|)!}\cdot \epsilon \cdot \frac{|Auto(p, \mathbf{ord})|}{|Auto(p, \emptyset)|}, \label{equation:matchestimation}
\end{align}
\end{small}
where the first term refers to the number of all possible assignments from $V(p)$ to $V(d)$, and $Auto(p, \mathbf{ord})$ refers to the automorphisms of $p$ that satisfy $\mathbf{ord}$. Please note that for the original pattern graph $p$, there is always only one automorphism satisfying $\mathbf{ord}$, because we apply SimB to break all other automorphisms of $p$ using $\mathbf{ord}$. However, the decomposed subgraphs of $p$ may have several automorphisms w.r.t.. $\mathbf{ord}$, so the last term is not a constant. Again, when $p$ has $O(1)$ vertices, the evaluation of Equation \ref{equation:matchestimation} can be evaluated in $O(1)$ time.

\textbf{Comparison.} Authors in \cite{lai2016scalable} also propose a function to estimate the match size. That function relaxes the degree constraints, that is, it allows a vertex in $p$ to be mapped to a vertex with smaller degree in $d$ with a small probability. However, in practice this probability should be $0$. Meanwhile, it doesn't consider the partial order $\mathbf{ord}$, which may filter out part of the matches. Thus, the result of that function includes some invalid matches. Our calculation considers there two factors and thus provides a smaller and more precise result.

\subsection{Revised Pattern Decomposition and CC-Join}
Because we use match compression in DDSL, the pattern decomposition and join process need to be adjusted to support compressed matches. In this section we describe how DDSL maintains the compressed form $f|s$ from the beginning, and keeps this form along the whole join process.

Given a pattern $p$ and a vertex cover $V_c(p)$, DDSL decomposes $p$ into a set of join units $Q=\{q_1, q_2,\ldots,q_k\}$, which satisfies the following conditions:
\begin{enumerate}
\item Each $q_i$ is an R1 unit;
\item $\bigcup\limits_{q_i \in Q} q_i = p$;
\item Each $q_i$'s anchor vertex is in $V_c(p)$.
\end{enumerate}
Condition (1) and (2) are common requirements in a join-based framework, to ensure that $M(q_i, d)$ can be directly listed and used to recover $M(p, d)$. We add condition (3) to make sure that the compressed form of matches of each $q_i$ w.r.t. $V_c(p)\cap V(q_i)$ can be listed directly from $\Phi(d)$.

Proposition \ref{proposition:subvertexcover} implies that $V_c(p)\cap V(q_i)$ is a vertex cover of $q_i$, which means we can compress the matches of $q_i$ w.r.t. $V_c(p)\cap V(q_i)$.

\begin{proposition}
\label{proposition:subvertexcover}
Given any graph $p$, a vertex cover $V_c(p)$, and its subgraph $p'$, $V_c(p)\cap V(p')$ is a vertex cover of $p'$.
\begin{proof}
For any edge $(v_i, v_j) \in E(p')$, without loss of generality, assume that $(v_i, v_j)$ is covered by $v_i \in V_c(p)$. It is obvious that $v_i$ is also in $V(p')$, so $v_i \in V_c(p) \cap V(p')$.
\end{proof}
\end{proposition}

Then we prove in Theorem \ref{theorem:listcompression} that the compressed matches of each $q_i$ w.r.t. $V_c(p)\cap V(p')$ can be listed directly from $\Phi(d)$. We also summarize the process in Algorithm \ref{algoritm:unitlisting}.

\begin{theorem}
\label{theorem:listcompression}
Given an R1 unit $q$ and a vertex cover $V_c(q)$, if $q$'s anchor vertex is in $V_c(q)$, then the compressed matches of $q$ w.r.t. $V_c(q)$ can be listed directly from $\Phi(d)$ without join.
\begin{proof}
For any skeleton $s:V_c(q) \mapsto V(d)$, the compressed match $f|s$ is built upon the conditional match set $M|s=\{f_i|s_i=s\}$. Let $v$ be the anchor vertex of $q$, and $s(v)=u_x$. By definition we know that $\forall f \in M|s, f(v)=u_x$, therefore $\forall d_j,h(x)\ne j:M|s \cap M_{ac}(q, d_j)=\emptyset$. Then we have $M|s \subseteq M(q, loc(u_x)) \subseteq M_{ac}(q, d_{h(x)})$. Since $M|s$ can be computed from $d_{h(x)}$, we can also obtain the compressed form $f|s$ from $d_{h(x)}$.
\end{proof}
\end{theorem}

\begin{algorithm}
    \caption{Unit Match Listing}
    \label{algoritm:unitlisting}
    \SetKwInOut{Input}{Input}
    \SetKwInOut{Output}{Output}
    \SetKwFunction{Compress}{Compress}
    \Input{$d_i$, $q$, $V_c(q)$, $q$'s anchor vertex $v$ which is in $V_c(q)$, $\mathbf{ord}$.}
    \Output{Compressed match set of $q$ in $d_i$.}
    \BlankLine
    $result \leftarrow \emptyset$\;
    \ForEach{$\; u \in d_i\text{'s center vertices} \;$}{
        $M_u \leftarrow \{f|f(v)=u\}$\;
        \ForEach{distinct $\; s \in M_u \; \text{w.r.t.} \; V_c(q) \;$}{
            $M|s \leftarrow \{ f_i|f_i \in M_u \; \text{and} \; s_i=s\}$ \;
            $f|s \leftarrow$ \Compress{$M|s$}\;
            Put $\;f|s\;$ into $\;result$\;
        }
    }
    \Return{$result$}\;
\end{algorithm}

After listing the compressed matches of each $q_i$ from $\Phi(d)$, the next step is to join the matches together. When matches are not compressed, the join of two matches $f_1:V(p_1) \mapsto V(d)$ and $f_2:V(p_2) \mapsto V(d)$ must satisfy that $\forall v \in V(p_1)\cap V(p_2): f_1(v)=f_2(v)$, and the join process is simply adding all mappings in $f_1$ and $f_2$ together, then check the validity of the result. In DDSL, all the matches are compressed, so the join condition and join process are modified. Before we describe the details, it is important to show the consistency of the compressed forms.

\begin{proposition}
\label{proposition:compressconsistency}
Given $p$, $V_c(p)$, $p$'s subgraph $p'$ and $v \in V(p')$, $v$ is a compressed vertex of $p'$ w.r.t. $V_c(p)\cap V(p')$ \textit{if and only if} $v$ is a compressed vertex of $p$ w.r.t. $V_c(p)$.
\begin{proof}
Since $v \in V(p')$, it is trivial that $v \notin V_c(p)\cap V(p')$ if and only if $v \notin V_c(p)$.
\end{proof}
\end{proposition}

Proposition \ref{proposition:compressconsistency} implies that a vertex $v$ will remain compressed (or not compressed) in every $f|s$, if we compress the matches of each subgraph $p'$ w.r.t. $V_c(p)\cap V(p')$. We call all such compressed matches the \textit{Consistently Compressed} (CC) matches under $V_c(p)$. Apparently in DDSL, the matches of join units are CC matches under $V_c(p)$. To join two CC match sets, we perform the following \textit{CC-join}.

\textbf{Join Condition.}
We reduce the condition on only the skeletons of two matches. Given $p$ and $V_c(p)$, two compressed matches $f_1|s_1$ of $p_1$ and $f_2|s_2$ of $p_2$ can be joined if $\forall v \in V_c(p)\cap V(p_1)\cap V(p_2): s_1(v)=s_2(v)$.

\textbf{Join Process.} During the join process, compressed and uncompressed vertices are handled separately. For uncompressed vertices, we simply union the mappings together. For a compressed vertex $v$, if $v$ appears in both two matches, say $f_1|s_1$ and $f_2|s_2$, then we take the intersection of two mapped sets, i.e., $v \mapsto f_1|s_1(v)\cap f_2|s_2(v)$, otherwise we just keep the only mapping of $v$. Suppose the join result is $f_3|s_3$, we then check the validity of each compressed vertex $v$, and remove $u$ from $f_3|s_3(v)$ if $v \mapsto u$ does not appear in any valid decompressed match. Finally we return $f_3|s_3(v)$ as the result.

\begin{algorithm}
    \caption{CC-Join}
    \label{algoritm:matchjoin}
    \SetKwInOut{Input}{Input}
    \SetKwInOut{Output}{Output}
    \SetKwFunction{Decompress}{Decompress}
    \Input{$V_c(p)$, $f_1|s_1$ of $p_1$, $f_2|s_2$ of $p_2$, and $\mathbf{ord}$.}
    \Output{Compressed match $f_3|s_3$.}
    \BlankLine
    $p_3 \leftarrow p_1 \cup p_2$; $\;\;\;s_3 \leftarrow s_1 \cup s_2$; $\;\;\;f_3 \leftarrow f_1|s_1$\;
    \ForEach{$\; v \in V(p_2)\backslash V_c(p) \;$}{
        \eIf{$\; v \in V(p_1) \;$}{
            $f_3(v) \leftarrow f_3(v) \cup f_2|s_2(v)$\;
        }{
            $f_3(v) \leftarrow f_2|s_2(v)$\;
        }
    }
    \ForEach{$\; v \in V(p_3) \backslash V_c(p) \;$}{
        \ForEach{$\; u \in f_3|s_3(v) \;$}{
            \If {$\; u \;$ is not valid}{
                Remove $u$ from $f_3|s_3(v)$\;
            }
        }
    }
    \Return{$f_3|s_3$}\;
\end{algorithm}

The join process is summarized in Algorithm \ref{algoritm:matchjoin}. Theorem \ref{theorem:ccjoincorrect} shows that 
after performing CC-join on two \textit{correct} CC matches, the result is also a \textit{correct} CC match. Therefore, we can gradually join the CC matches of join units (which are correct) together in a specific order, to obtain all the \textit{correct} CC matches of the pattern graph $p$.




\begin{theorem}
\label{theorem:ccjoincorrect}
Supposing that by joining two CC matches $f_1|s_1$ of $p_1$ and $f_2|s_2$ of $p_2$, we get the match $f_3|s_3$ of $p_3$, then $f_3|s_3$ is a correct CC match of $p_3$ under $V_c(p)$, which has skeleton $s_3$.
\begin{proof}
We first prove the consistency, i.e., $f_3|s_3$ is a CC match under $V_c(p)$, which has skeleton $s_3$.

During the join process, we compute $s_3$ as $s_1 \cup s_2$, thus $\forall v \in V(p_3)$, $v$ is a uncompressed vertex if and only if $v \in V_c(p)$. Similarly, $\forall v \in V(p_3)$, $v$ is a compressed vertex if and only if $v \notin V_c(p)$.

Then we prove the correctness. The skeleton $s_3$ is obtained by a regular join of two uncompressed matches, so its correctness is guaranteed, and we only focus on compressed vertices.
\begin{enumerate}
\item For any $f \in M(p_3, d)$, which has skeleton $s_3$, we examine an arbitrary vertex $v \in V(p_3)\backslash V_c(p)$. Since $v$ is either in one of $V(p_1)\backslash V_c(p)$ and $V(p_2)\backslash V_c(p)$, or in both, we discuss these two situations separately:
    \begin{itemize}
    \item $v$ is in either $V(p_1)\backslash V_c(p)$ or $V(p_2)\backslash V_c(p)$. Without loss of generality, assume that $v \in V(p_1)\backslash V_c(p)$. Let $f'=\{v_i \mapsto f(v_i)|\forall v_i \in v_1\}$, it is clear that $f' \in M(p_1, d)$. Since $f_1|s_1$ is correct, $f'(v)=f(v)$ must be in $f_1|s_1(v)$, otherwise $f'$ will be lost when decompressing $f_1|s_1$. Thus $f(v)$ is in $f_3|s_3(v)$.
    \item $v$ is in both $V(p_1)\backslash V_c(p)$ and $V(p_2)\backslash V_c(p)$. Let $f'=\{v_i \mapsto f(v_i)|\forall v_i \in v_1\}$, and $f''=\{v_i \mapsto f(v_i)|\forall v_i \in v_2\}$. Similar to the discussion above, we have $f'(v) \in f_1|s_1(v)$ and $f''(v) \in f_2|s_2(v)$, and thus $f(v)$ is in $f_1|s_1(v) \cap f_2|s_2(v)$, which is exactly $f_3|s_3(v)$.
    \end{itemize}
Thus, for each $v \in V(p_3)\backslash V_c(p)$, and for each $f \in M(p_3, d)$, we have $f(v) \in f_3|s_3(v)$. When decompressing $f_3|s_3$, any valid $f \in M(p_3, d)$ will appear in the Cartesian product, and be in the output.
\item The analysis in (1) also holds in the opposite direction, that is, given an arbitrary vertex $v \in V(p_3)\backslash V_c(p)$ and any $u \in f_3|s_3(v)$, $v \mapsto u$ exists either in valid matches in $M(p_1, d)$ (or $M(p_2, d)$), or in both. In each case we can derive that $\forall (v, v') \in E(p_3): (u, f_3|s_3(v')) \in E(d)$. Thus, each decompressed match preserves all edges in $p_3$, and is a valid match if it passes the validation check in Section \ref{section:decompression}.
\end{enumerate}
\end{proof}
\end{theorem}


\subsection{Revised Optimal Compression}
In general, a join process is parallelized in the following way (supposing we are joining $M_1|s_1$ of $p_1$ and $M_2|s_2$ of $p_2$).
\begin{enumerate}
    \item We know the join condition is built on $V_{key}=V(p_1) \cap V(p_2) \cap V_c(p)$, and thus we can generate a join key $\{s(v)|\forall v \in  V_{key}\}$ for each $f|s$ involved in this join. It is obvious that any two matches satisfying the join condition must have the same join key;
    \item The distributed system can aggregate matches with the same key together in a worker, and then different workers can perform the join on different join keys simultaneously.
\end{enumerate}

If we use the optimal compression which maximizes $R_{lower}$, it may not be suitable for parallelizing the join process in a distributed system. For example, in Fig. \ref{fig:join}, the optimal compression is the one in (c), however, using this compression we cannot generate the join key. Thus, the join degenerates to the Cartesian product between all matches and cannot be parallelized efficiently. Specifically, Lemma \ref{lemma:paralleljoin} shows that to well parallelize the join process, the induced graph $p[V_c(p)]$ should be connected.

\begin{lemma}
\label{lemma:paralleljoin}
Given $p$ and $V_c(p)$, if $p[V_c(p)]$ has at least two connected components, then during the CC-join process, there is at least one CC-join involving two patterns $p_1$ and $p_2$, such that $V(p_1) \cap V(p_2) \cap V_c(p)=\emptyset$.
\begin{proof}
Treat the whole process as a join tree. The root corresponds to $p$ it self, where $p[V(p)\cap V_c(p)]$ has more than one connected component. Each leaf corresponds to a join unit $q$, where $q[V(q)\cap V_c(p)]$ has only one connected component, because $q$'s anchor vertex is in $V_c(p)$, and it is connected to all other vertices in $V(q)$. Therefore, we can find at least one node in the join tree, such that
\begin{enumerate}
\item it corresponds to a pattern $p'$, where $p'[V(p')\cap V_c(p)]$ has two connected components, and
\item its children corresponds to pattern $p_1$ and $p_2$, where $p_1[V(p_1)\cap V_c(p)]$ and $p_2[V(p_2)\cap V_c(p)]$ both have only one connected component.
\end{enumerate}
In this case, we have
\begin{align*}
p'[V(p')\cap V_c(p)] &= p'[(V(p_1) \cup V(p_2))\cap V_c(p)]  \\
                     &= p'[(V(p_1)\cap V_c(p)) \cup (V(p_2) \cap V_c(p))].
\end{align*}
We know that $p'[V(p_1)\cap V_c(p)]=p_1[V(p_1)\cap V_c(p)]$ and $p'[V(p_2)\cap V_c(p)]=p_2[V(p_2)\cap V_c(p)]$, both of which are connected. Thus, the two connected components in $p'[V(p')\cap V_c(p)]$ are exactly $p'[V(p_1)\cap V_c(p)]$ and $p'[V(p_2)\cap V_c(p)]$, which implies $V(p_1) \cap V(p_2) \cap V_c(p)=\emptyset$.
\end{proof}
\end{lemma}

To facilitate the join process, we require the induced graph $p[V_c(p)]$ to be connected. Under this condition, we pick the compression that maximizes $R_{lower}$. Such a compression is called an \textit{optimal connected compression}.

\section{The Optimal Join Tree}
\label{section:optimaltree}
Given a set of join units $Q$ and a specific join order, the whole join process forms a join tree. A typical join tree $\mathcal{T}(p)$ contains a set $Q(\mathcal{T}(p))=\{q_1, \ldots, q_k\}$, and a set $P(\mathcal{T}(p))=\{p_1, p_2, \ldots, p_l\}$, where each $q_i$ is a join unit on a leaf node, and each $p_i$ is a subgraph of $p$ on an internal node. Assuming that we already have the optimal connected compression of $p$, choosing different join trees may give us different performance, depending on how much intermediate result is produced. Thus, in DDSL we try to find the optimal join tree, aiming at minimizing the I/O and communication cost produced during the whole join process.

\begin{definition}[Optimal Join Tree]
Given $p$, $V_c(p)$, and a cost function $Cost$, a join tree $\mathcal{T}(p)$ is optimal if
\begin{enumerate}
    \item The root of $\mathcal{T}(p)$ corresponds to $p$.
    \item Each leaf in $\mathcal{T}(p)$ corresponds to a join unit.
    \item It minimizes $Cost(\mathcal{T}(p))$ among all possible trees that satisfy condition (1) and (2).
\end{enumerate}
\end{definition}

In MapReduce, the I/O cost and communication cost come from the map data, the shuffle data, and the reduce data. Following the join process, we compute the total cost of any specific join tree $\mathcal{T}(p)$ by adding up the cost from each part.

\textbf{Cost of Processing Leaves}. To list all matches of a join unit $q$, the mappers take $\Phi(d)$ as the input, and output all compressed matches of $q$. The output is directly writing into disk, so no shuffle or reduce is needed. Suppose storing $\Phi(d)$ and all matches of $q$ requires $S(\Phi(d))$ and $S(q)$ storage, respectively, then the cost is $S(\Phi(d)) + S(q)$. In DDSL, we list the matches of all join units in a single MapReduce round, and for each unit we store its matches in a unique file. In this way, the total cost for processing all leaf nodes in $\mathcal{T}(p)$ is
\begin{small}
\begin{align*}
S(\Phi(d)) + \sum\limits_{q \in Q(\mathcal{T}(p))}S(q).
\end{align*}
\end{small}

\textbf{Cost of Processing Internal Nodes}. To list all matches of a non-unit pattern $p_i=p_i^l \cup p_i^r$, the mappers take the matches of $p_i^l$ and $p_i^r$, and output each match associated with the join key. All key-match pairs are then read and shuffled to their corresponding reducers. Finally each reducer takes matches with the same join key, and output the join result. The total cost incurred in map, shuffle and reduce is $2\cdot S(p_i^l) + 2\cdot S(p_i^r)$, $2\cdot S(p_i^l) + 2\cdot S(p_i^r)$ and $S(p_i^l) + S(p_i^r) + S(p_i)$, respectively.

\textbf{Cost of Decompression}. After we computing the compressed matches of $p$, we may need to decompress the results to get the plain matches. This can be done in a single map function. The mappers take $S(p)$ as the input, and output the decompressed matches. Therefore the cost is simply $S(p) + |V(p)|\cdot |M(p, d)|$. Note that, this decompression is optional. We can decompress the result when we need the plain form, or keep it the compressed form otherwise.

By summarizing the cost of processing all nodes in $\mathcal{T}(p)$ and rearranging the terms associated with each pattern, we write the total cost of $\mathcal{T}(p)$ as
\begin{small}
\begin{align}
&Cost\big(\mathcal{T}(p)\big) = \nonumber \\
&\sum_{p_i \in P \cup Q}6\cdot S(p_i) + S\big(\Phi(d)\big) + 2\cdot S(p) + |V(p)|\cdot |M(p, d)|. \label{equation:treecost}
\end{align}
\end{small}

The cost $S\big(\Phi(d)\big)$ and $S(p) + |V(p)|\cdot |M(p, d)|$ introduced in processing the leaf nodes and the final decompression, are constants if $p$ and $d$ do not change, so we ignore these terms when possible, and write Equation \ref{equation:treecost} in a recursive form:
\begin{small}
\begin{align*}
&Cost\big( \mathcal{T}(p_i) \big) = \\
& \begin{cases}
      S(p_i), & \text{if } p_i \text{ is a join unit} \\
      \\
      Cost\big( \mathcal{T}(p_i^l) \big) + Cost\big( \mathcal{T}(p_i^r) \big) \\
      \ \ \ + 5\cdot S(p_i^l) + 5\cdot S(p_i^r) + S(p_i), & \text{otherwise}.
  \end{cases}
\end{align*}
\end{small}

It suggests that we can use dynamic programming to compute the optimal join tree, which is summarized in Algorithm \ref{algorithm:jointree}. 
In line $1-3$ we initialize the tree set $T$ with all valid join units. From line $4$ to line $10$ we gradually union the small patters to bigger ones, and keep record of the minimum cost. Note that line $6$ ensures that the join key can be generated for every possible join. In line $11$ we recursively construct the whole join tree by following its children all the way to the join units.

\begin{algorithm}
    \caption{Find Optimal Join Tree}
    \label{algorithm:jointree}
    \SetKwInOut{Input}{Input}
    \SetKwInOut{Output}{Output}
    \SetKwFunction{Cost}{Cost}
    \Input{$V_c(p)$, $p$.}
    \Output{The optimal join tree $\mathcal{T}_{opt}(p)$.}
    \BlankLine
    $T \leftarrow \emptyset\;$\;
    \ForEach{valid join unit $q_i$ of $\;p\;$}{
        $T(q_i) \leftarrow (\Cost{$q_i$}, null, null)$\;
    }
    \For{$i \leftarrow 1$ \emph{\KwTo} $|E(p)|$}{
        \ForEach{$p_i \in T \text{ s.t. } |E(p_i)|=i$}{
            \ForEach{$p_j \in T \text{ s.t. } V(p_i)\cap V(p_j)\cap V_c(p) \ne \emptyset$}{
                $p' = p_i \cup p_j$\;
                compute $\Cost{$p'$}$\;
                \If{$p' \notin T$ \bf{or} $T(p').cost > \Cost{$p'$}$}{
                    $T(p') \leftarrow (\Cost{$p'$}, p_i, p_j)$\;
                }
            }
        }
    }
    \Return{$\mathcal{T}_{opt}(p)$ constructed from $T(p)$}\;
\end{algorithm}

Lemma \ref{lemma:opttree} shows that Algorithm \ref{algorithm:jointree} can find the join tree with minimum estimation cost. Since there can be at most $2^{|E(p)|}$ different subgraphs of $p$, and each of them might be joined with any other patterns, the time complexity of this algorithm is $O(4^{|E(p)|})$. Considering that we filtered out many invalid joins, the actual running time is much smaller in practice.

\begin{lemma}
\label{lemma:opttree}
The join tree found by Algorithm \ref{algorithm:jointree} is optimal (has the lowest estimation cost).
\begin{proof}
We use the strong induction to prove that every node in the optimal join tree is stored in $T$.
\begin{enumerate}
    \item For each leave node $q_i$, its processing cost is directly computed and stored in $T$ (lines 2-3);
    \item For each non-leave node $p' = p_i \cup p_j$, if $p_i$ and $p_j$ are in $T$, they will be enumerated in lines 5-6, so node $p'$ will be explored and stored in $T$ (lines 8-10).
    \item For each non-leave node $p' = p_i \cup p_j$, assuming that $|V(p')|>|V(p_i)|\geq |V(p_j)|$, line 4 and 5 guarantee that nodes $p_i$ and $p_j$ are explored and stored in previous steps.
\end{enumerate}

Therefore, the root node $T(p)$ of the optimal tree is in $T$, and is returned at line 11.
\end{proof}
\end{lemma}

\section{Incremental Updating}
\label{section:incremental}
In this section we address the problem of handling dynamic graphs. Assuming that a batch $U$ of edge updates is applied on the original data graph $d$, resulting in a new data graph $d'$, we update $\Phi(d)$ and $M(p, d)$ in three steps:
\begin{enumerate}
\item Update $\Phi(d)$ to $\Phi(d')$ according to $U$;
\item Compute a patch set $M_{new}(p, d')$ containing all newly-appeared matches, i.e., $M_{new}(p, d') = M(p, d')\backslash M(p, d)$;
\item Compute $M(p, d')$ by first filtering matches in $M(p, d)$ that should be removed, and then merging the result with $M_{new}(p, d')$.
\end{enumerate}

\subsection{Update the NP Storage}

To update the NP storage $\Phi(d)$, essentially we need to update each local graph $loc(u_i)$ correctly. We first discuss how an edge insertion/deletion influences the edges in $loc(u_i)$, and then extend the discussion to $d_j$, where multiple local graph exists.

\textbf{Edge Insertion}. There are two possible situations that inserting an edge can lead to the change of $loc(u_i)$.
\begin{enumerate}
\item Inserting edge $(u_i, u_j)$, which means $u_j$ becomes a new neighbor of $u_i$. In this case, we need to check all edges adjacent to $u_j$, and add edge $(u_j, u_k)$ into $loc(u_i)$ if $u_k$ is $u_i$'s neighbor.
\item Inserting edge $(u_j, u_k)$, where $u_j$ and $u_k$ are both $u_i$'s neighbors. In this case, we only need to add $(u_j, u_k)$ into $loc(u_i)$.
\end{enumerate}

\textbf{Edge Deletion}. Edge deletion is the reverse of edge insertion, so it can be handled by the above two situations with all edge insertions replaced with edge deletions. The only difference is that, in situation (1), we can just delete all edges in $loc(u_i)$ which are adjacent to $u_j$.

Extending our discussion to the update of $d_k$, there are in total three possible cases.
\begin{enumerate}[C$_1$]
\item Inserting/deleting edge $(u_i, u_j)$, where $u_i$ and $u_j$ are both center vertices of $d_k$. All edges adjacent to $u_i$ or $u_j$ must be preserved to store $loc(u_i)$ and $loc(u_j)$, respectively, so we only need to insert/delete edge $(u_i, u_j)$.
\item Inserting/deleting edge $(u_i, u_j)$, where $u_i$ and $u_j$ are both border vertices of $d_k$. For all local graphs in $d_k$, it belongs to situation (2) of insertion/deletion, so we only need to insert/delete edge $(u_i, u_j)$.
\item Inserting/deleting edge $(u_i, u_j)$, where $u_i$ is a center vertex, and $u_j$ is a border vertex. For insertion, we deal as situation (1). For deletion, we first delete $(u_i, u_j)$, then for each edge adjacent to $u_j$, we delete it only if no other local graphs contain it.
\end{enumerate}

In DDSL we deal with all three cases in a single MapReduce round.
Basically, when inserting an edge $(u_i, u_j) \text{ s.t. } h(i)\ne h(j)$, we need to deal with case C$_3$ in both $d_{h(i)}$ and $d_{h(j)}$, which requires to know $\neighbor[d']{u_j}$ and $\neighbor[d']{u_j}$, respectively. Thus, we use the map function to prepare such data, and use the reduce function to update $\Phi(d)$. Note that the reduce function takes $\mathcal{N}_{d_k'}$ as one of the inputs, which contains all neighbor sets necessary for adding edges into $d_k$. When dealing with case C$_3$, the reduce function can get $\neighbor[d']{u_i}$ (or $\neighbor[d']{u_j}$) from $\mathcal{N}_{d_k'}$. The whole algorithm is summarized in Algorithm \ref{algorithm:updatestorage}.

\begin{algorithm}[h!]
    \caption{Update NP storage}
    \label{algorithm:updatestorage}
    \SetKwInOut{Input}{Input}
    \SetKwInOut{Output}{Output}
    \Input{$\Phi(d)$, $h$, $U$.}
    \Output{$\Phi(d')$.}
    \SetKwFunction{output}{Output}
    \SetKwFunction{curN}{NeighborSet}
    \SetKwProg{Function}{Function}{ }{end}
    \BlankLine
    \Function{NeighborSet($d_k$, $i$, $U$)\tcp*[f]{compute $\neighbor[d']{u_i}$}}{
        return \parbox[t]{.6\linewidth}{%
    $\neighbor[d_k]{u_i} \; \backslash \; \{u_l|\forall (u_i, u_l) \in E_d(U)\}$ \\
    $\cup \; \{u_l|\forall (u_i, u_l) \in E_a(U)\}$\;}

    }
    \Function{map($d_k$, $h$, $U$)}{
        \ForEach{$\; (u_i, u_j) \in E_a(U) \;$}{
            \If{$\; h(i)=k$ \bf{and} $h(i) \ne h(j) \;$}{
                \output{$h(j), \curN{$d_k$, $i$, $U$}$}\;
            }
            \If{$\; h(j)=k$ \bf{and} $h(i) \ne h(j) \;$}{
                \output{$h(i), \curN{$d_k$, $j$, $U$}$}\;
            }
        }
    }
    \Function{reduce($d_k$, $h$, $U$, $\mathcal{N}_{d_k'}$)}{
        \ForEach{$(u_i, u_j) \in E_d(U)$}{
            delete edge $(u_i, u_j)$ from $d_k$\;
        }
        \ForEach{$(u_i, u_j) \in E_a(U)$}{
            insert edge $(u_i, u_j)$ into $d_k$ using $\mathcal{N}_{d_k'}$\;
        }
        \output{updated $d_k$}\;
    }
\end{algorithm}

Each mapper takes $d_k$ and $U$ as the input, which causes $S(\Phi(d)) + m\cdot |E(U)|$ I/O cost. All mappers output the neighbor sets of vertices in $U$, which are shuffled to the reducer. This incurs at most $3\cdot \sum_{u_i \in U}|\neighbor[d']{u_i}|$ communication cost. The reducers read $d_k$, $U$ and $\mathcal{N}_{d_k'}$, and then output $\Phi(d')$, so the I/O cost of reduce is $S(\Phi(d)) + m\cdot |E(U)| + \sum_{u_i \in U}|\neighbor[d']{u_i}| + S(\Phi(d'))$. In summary, the total cost of updating $\Phi(d)$ is
\begin{align*}
2\cdot S(\Phi(d)) + 2\cdot m\cdot |E(U)| + 4\cdot \sum_{u_i \in U}|\neighbor[d']{u_i}| + S(\Phi(d'))
\end{align*}

\subsection{Update the Match Set}
The update of the match set can be divided into two categories: the removed matches and the newly-appeared matches. In this section we discuss how to identify or extract them separately.

The removed matches exist in $M(p, d)$, but are no longer valid because some edges are deleted. Lemma \ref{lemma:removematch} shows that removed matches can be easily detected by checking each match with the deleted edge set $E_d(U)$. Thus, we can remove those matches from $M(p, d)$ during the decompression step, and only output matches that should not be removed. The extra I/O cost comes from reading the deleted edge set $E_d(U)$ by each mapper, which is in total $m\cdot |E(U)|$.

\begin{lemma}
\label{lemma:removematch}
A match should be removed from $M(p, d)$ if and only if it maps any edge in $E(p)$ to a deleted edge in $E_d(U)$.
\begin{proof}
\textit{(If).} It is trivial according to the definition of a match.

\textit{(Only if).} Suppose $f \in  M(p, d)$ and $f \notin M(p, d')$. Since $f \notin M(p, d')$, there must be an edge $(v_i, v_j) \in E(p)$ s.t. $(f(v_i), f(v_j)) \notin E(d')$. Since $f \in M(p, d)$, $(f(v_i), f(v_j)) \in E(d)$. Thus edge $(f(v_i), f(v_j))$ is deleted from $E(d)$.
\end{proof}
\end{lemma}

The newly-appeared matches do not exist in $M(p, d)$, but appear in $M(p, d')$. Lemma \ref{lemma:newmatch} shows that a newly-appeared match must map at least one edge in $E(p)$ to an inserted edge in $E_a(U)$. Therefore we can use $E_a(U)$ to identify these matches, which consist the patch set $M_{new}(p, d')$.

\begin{lemma}
\label{lemma:newmatch}
A match in $M(p, d')$ is not in $M(p, d)$ if and only if it maps any edge in $E(p)$ to an inserted edge in $E_a(U)$.
\begin{proof}
\textit{(If).} Suppose that $f \in M(p, d')$ maps $(v_i, v_j) \in E(p)$ to $(u_i, u_j) \in E_a(U)$. Since $(u_i, u_j) \notin E(d)$, $f \notin M(p, d)$.

\textit{(Only if).} Suppose $f \in  M(p, d')$ and $f \notin M(p, d)$. Since $f \notin M(p, d)$, there must be an edge $(v_i, v_j) \in E(p)$ s.t. $(f(v_i), f(v_j)) \notin E(d)$. Since $f \in M(p, d')$, $(f(v_i), f(v_j)) \in E(d')$. Thus edge $(f(v_i), f(v_j))$ is inserted into $E(d)$.
\end{proof}
\end{lemma}

Recall that in the join-based framework, we decompose the pattern $p$ into join units. It means for each edge $(v_i, v_j) \in E(p)$, we can find at least one join unit $q_i$, s.t. $(v_i, v_j) \in E(q_i)$. By forcing all matches of $q_i$ to map $(v_i, v_j)$ to an inserted edge in $E_a(U)$, we can ensure that all join results also map $(v_i, v_j)$ to an inserted edge, and thus are newly-appeared matches. Based on this, we compute the patch set $M_{new}(p, d')$ in three steps:
\begin{enumerate}
\item Decompose $p$ into a set $Q$ of join units.
\item For each join unit $q_i \in Q$, compute a partial patch set $M_{new}(p, d', q_i)$, which contains all matches in $M(p, d')$ that map at least one edge in $E(q_i)$ to an inserted edge.
\item Compute $M_{new}(p, d') = \bigcup\limits_{q_i \in Q}M_{new}(p, d', q_i)$.
\end{enumerate}

A straightforward way of doing this is using the bushy-join described before. However, it may not save the I/O cost. In fact, Example \ref{example:bushycost} shows that in a regular join process, computing $M(p, d')$ using $M_{new}(p, d')$ may incur more I/O cost than directly computing $M(p, d')$.

\begin{example}
\label{example:bushycost}
Suppose we have a join $p=q^l \cup q^r$. As discussed, directly computing $M(p, d')$ with this join requires listing and joining $M(q^l, d')$ and $M(q^r, d')$. If we want to compute $M(p, d')$ using $M_{new}(p, d')$, we need to first compute $M_{new}(p, d', q^l)$ by listing and joining $M_{new}(q^l, d', q^l)$ with $M(q^r, d')$, and then compute $M_{new}(p, d', q^r)$ similarly. Finally we compute $M(p, d')$ by merging $M(p, d)$, $M_{new}(p, d', q^l)$ and $M_{new}(p, d', q^r)$, with invalid matches filtered out. In comparison with direct computing, we pay extra I/O cost for reading/writing $M(p, d)$, $M_{new}(q^l, d', q^l)$ and $M_{new}(q^r, d', q^r)$, and do not save any I/O cost.
\end{example}

In DDSL, we design a \textit{navigated join (Nav-join)} to compute $M_{new}(p, d')$ with a lower cost based on two assumptions:
\begin{itemize}
\item The inserted edge number is small compared to $E(d)$, so the size of $M_{new}(p_i, d', q_i)$ for any $p_i$ and $q_i$ should be small.
\item The size of $M(p_i, d)$ for a small $p_i$ is usually much larger than $|E(d)|$. 
\end{itemize}
If these two assumptions hold, then instead of joining two match sets, we may use a partition-and-expand strategy to reduce the join cost. Basically, for a join $p_i=p_j \cup q_k$, where $q_k$ is a join unit, we can send each $f \in M_{new}(p_j, d')$ to several partitions. Then inside each partition $d_x'$, we expand $f$ to get the matches of $p_i$. If we choose the partitions carefully, we can guarantee the correctness of the result. Suppose that $M_{new}(p_j, d')$ takes $S_{new}(p_j)$ storage, the total cost is now at most $S\big(\Phi(d')\big) + (4m+1)\cdot S_{new}(p_j)+S_{new}(p_i)$. According to our assumptions, this cost should be lower than the original bushy-join cost. Based on this idea, we design the Nav-join to compute $M_{new}(p, d', q_i)$ as follows:
\begin{enumerate}
\item Find a left-deep tree w.r.t. the join unit set $Q$, where $q_i$ is the lowest leaf.
\item Extract $M_{new}(q_i, d', q_i)$ using Algorithm \ref{algoritm:unitlisting} with the constraint that every match must map at least one edge in $q_i$ to an inserted edge.
\item For a join in the tree, we compute the result using Nav-join by partitioning the matches on the left side, and expanding them in each part of the NP storage.
\item Repeat step (3) from the bottom of the tree to the root, and the final result is $M_{new}(p, d', q_i)$.
\end{enumerate}
Note that all matches are still compressed w.r.t. a vertex cover of $p$. Example \ref{example:navjoin} demonstrates the process of a Nav-join with $3$ units and $2$ partitions. We now discuss the details and optimizations below.

\begin{figure}[htbp]
\centering
\includegraphics[width=.9\linewidth, keepaspectratio]{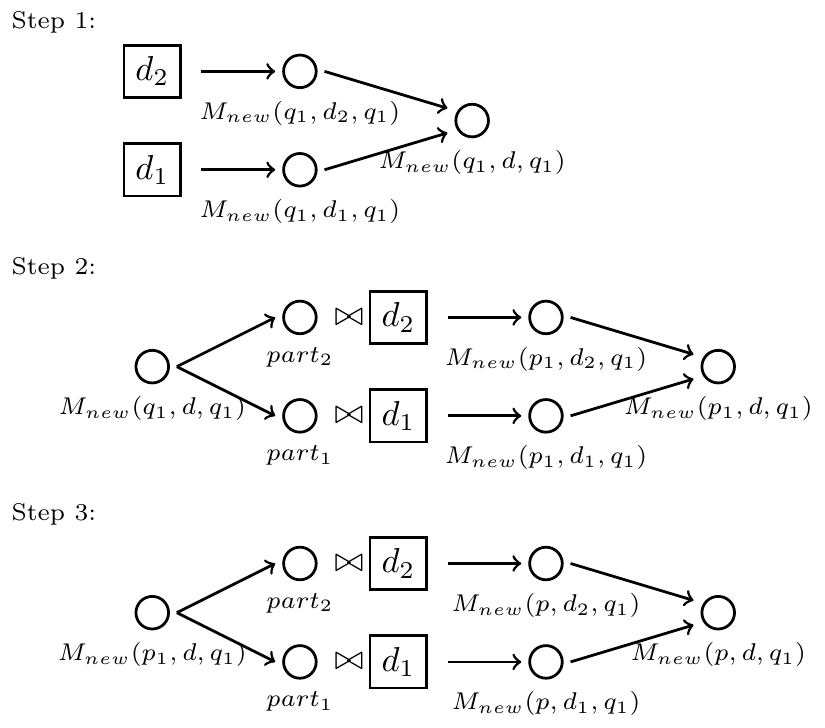}
\caption{An example of the Nav-join process.}
\label{fig:navjoin}
\end{figure}

\begin{example}
\label{example:navjoin}
Consider a join $p=(q_1 \cup q_2) \cup q_3$, and an NP storage of $2$ partitions. Fig. \ref{fig:navjoin} shows the $3$ steps to compute $M_{new}(p, d, q_1)$ using Nav-join. In the first step, we extract the new match sets of $q_1$ from two partitions separately, and then merge them together. In the second step, we first partition the previous result into $2$ parts, s.t. a match $f$ is in partition $i$ only if $f$ can be joined with some matches in $d_i$. Then inside each partition, we expand each match of $q_1$ to get the new matches of $p_1$, where $p_1=q_1 \cup q_2$. Similarly, in the third step, we partition and expand the matches of $p_1$ to get the match set $M_{new}(p, d, q_1)$.
\end{example}

\textbf{Optimal Left-Deep Tree.} In a join $p_i = p_i^l \cup p_i^r$, the mappers take $M_{new}(p_i^l, d', q_i)$ as input, and output the matches sent to each $d_k'$, which are shuffled to the reducers. The reducers take $\Phi(d')$ and received matches as input, and output $M_{new}(p_i, d', q_i)$. According to assumption (1), the main cost would be reading $\Phi(d')$, i.e., $S(\Phi(d'))$. Assuming a left-deep tree involves $j$ join units, then the total cost is $j\cdot S(\Phi(d'))$, which comes from $1$ unit match listing and $j-1$ Nav-joins. Thus, the optimal left-deep tree is the one involves the minimum number of join units.

\textbf{Match Navigation.} In step (3), supposing the join is $p_i = p_i^l \cup p_i^r$, we decide whether $f$ should be sent to $d_k'$ as follows:
\begin{enumerate}
\item If $p_i^r$'s anchor vertex $v$ is used to generate the join-key, and $f(v)=u_j$, then $f$ can only be joined with matches in $M(p_i^r, loc'(u_j))$, where $loc'(u_j)$ is the local graph of $u_j$ in $d'$. In this case, we only send $f$ to $d_{h(j)}'$.
\item Otherwise, we generate the join-key of $f$. If a match $f'$ can be joined with $f$, $f'$ must have the same join-key, and thus $f'(v)$ must be the common neighbors of all vertices in the join-key. For each vertex $u_j$ in the join-key, we send $f$ to $d_{h(j)}'$.
\end{enumerate}
Computing the common neighbors requires to know all neighbors of those vertices in the join-key, however, storing the neighbors of all vertices within $d_k'$ is costly. For example, in one of the datasets used in our experiments, we find a vertex with more than $10^5$ neighbors. If $m\approx10^2$, this vertex appears in almost every part of $\Phi(d)$, and it takes more than $10^7$ integers to store its neighbors in every partition. To reduce this storage, we use a bit map of length $m$ for each vertex $u$, to store the partition ids $h(l),\forall u_l \in \neighbor[d']{u}$, and check the common partition ids instead of common neighbors. If we use 32-bit integers to simulate bit maps, it takes $|V(d')|\cdot \lceil\frac{m}{32}\rceil$ integers in total. When performing the joins of $f$ in $d_k'$, we also use the join-key to filter out impossible local graphs in $d_k'$. Since the common neighbors now are available in $d_k'$, this filtering requires no other storages.

\textbf{Parallelize All Trees.} For each $q_i \in Q$ we need to compute $M_{new}(p, d', q_i)$ according to its left-deep tree. If we compute all sets in serialization, the total cost is $S(\Phi(d'))\cdot |Q|^2$. We noticed that in each join and unit match listing, the input $\Phi(d')$ can be shared, so we compute all match sets in parallelism. In each MapReduce round, we process the same level of all left-deep trees simultaneously. Since all left-deep tress has the same height $|Q|-1$, the calculation can be done in exact $|Q|$ rounds, and the total cost is thus $S(\Phi(d'))\cdot |Q|$.

\textbf{Match Deduplication.} For $M_{new}(p, d', q_i)$, there might be a match $f$, which maps an edge in $q_i$ to an inserted edge, and another edge in $q_j$ to an inserted edge. This $f$ will also appear in $M_{new}(p, d', q_j)$ according to our algorithm. To avoid such duplications, we assign a total order $\{q_i < q_j if i < j\}$ on $Q$, then $\forall f \in M_{new}(p, d', q_i)$, we keep it only if for any $q_j < q_i$, $f$ does not map an edge in $q_j$ to an inserted edge. In other words, $f$ can map an edge in $q_j$ to an inserted edge if and only if $q_j > q_i$.
\begin{theorem}
\label{theorem:dedupcorrect}
By assigning the total order in the Nav-join, there will be no duplicate matches or lost matches.
\begin{proof}
We first show that no math is lost. In the total order, we can find a smallest $q_i$, s.t. $q_i$ contains an edge $(v_i, v_j)$ which is mapped to an inserted edge by $f$. It is easy to verify that $f \in M_{new}(p, d', q_i)$.

We then show that there is no duplicate matches, i.e., if $i \ne j$, $M_{new}(p, d', q_i)\cap M_{new}(p, d', q_j)=\emptyset$. For any $f\in M_{new}(p, d', q_i)$, there are two possibilities.
\begin{enumerate}
\item If $f$ does not map an edge in $q_j$ to an inserted edge, then $f \notin M_{new}(p, d', q_j)$.
\item If $f$ maps an edge in $q_j$ to an inserted edge, then $i < j$, and thus $f \notin M_{new}(p, d', q_j)$.
\end{enumerate}
\end{proof}
\end{theorem}


\section{Experiments}
\label{section:experiments}
In this section, we conduct two parts of experiments to evaluate the cost of DDSL on static and dynamic graphs separately. Specifically, our experimental study has three main goals:
\begin{itemize}
    \item We illustrate the cheap construction cost and space cost of the NP storage mechanism.
    \item For static graphs, we compare the overall cost of DDSL to the state-of-the-art distributed methods.
    \item For dynamic graphs, we show the efficiency of DDSL in updating the NP storage as well as the matching result.
\end{itemize}

\subsection{Experiment Setup}
\subsubsection{Datasets and Queries.}
\begin{figure}[ht]
\centering
\includegraphics[width=\linewidth, keepaspectratio]{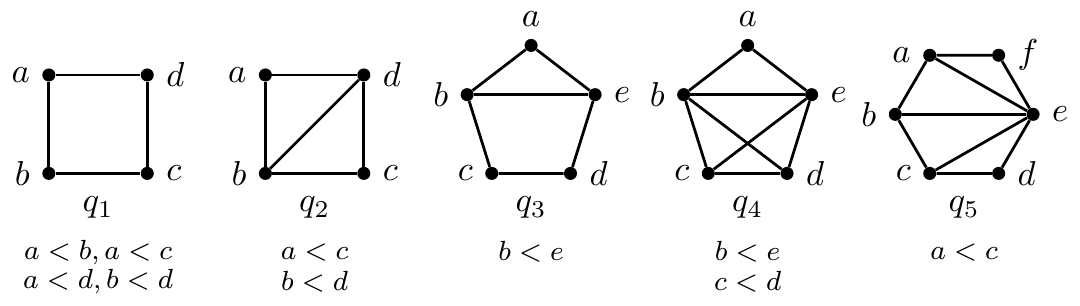}
\caption{Pattern graphs.}
\label{fig:patterns}
\end{figure}

In our experiments, we use $4$ real-word graphs WebGoogle(WG), WikiTalk(WT), LiveJournal(LJ), and UK-2002(UK) as the data graph, which that are commonly used in recent works \cite{lai2016scalable,kim2016dualsim,qiao2017subgraph}. Datasets WG, WT and LJ can be downloaded from SNAP \footnote{http://snap.stanford.edu/data/index.html}, and dataset UK can be downloaded from WEB \footnote{http://law.di.unimi.it}. The size of each dataset is listed in Table \ref{table:datasets}. For the pattern graphs, we pick $5$ commonly used ones from recent works \cite{lai2016scalable,kim2016dualsim,qiao2017subgraph}, which are shown in Fig. \ref{fig:patterns}. Below each pattern, we show the corresponding partial order $\mathbf{ord}$ used by SimB.

\begin{table}
\centering
\caption{Sizes of Datasets.}
\label{table:datasets}
\begin{tabular}{|c|c|c|c|c|}
\hline
 & WG & WT & LJ & UK                    \\
\hline
$|V|$ & 0.87M & 2.39M & 4.84M & 18.5M   \\
\hline
$|E|$ & 5.1M  & 5.0M  & 34M   & 227.5M  \\
\hline
\end{tabular}
\end{table}

\subsubsection{Compared Methods.}
In the experiments on static graphs, we compare DDSL with two state-of-the-art distributed approaches SEED \cite{lai2016scalable} and Crystal \cite{qiao2017subgraph}. These two methods are originally designed for the subgraph enumeration task, which does not write the result into disks. In our experiments we make them write all matches into disks to solve the subgraph listing problem.

In the experiments on dynamic graphs, since no other works target on the exact problem discussed in this paper, we use Delta-BigJoin \cite{ammar2018distributed}, a recent approach that handles directed and dynamic graphs, as the competitor. It should be noted that, the running time of DDSL and Delta-BigJoin cannot be directly compared, because these two methods are designed for different tasks. Even if we simulate undirected graphs by duplicating edges, it is still unfair, since some of the optimizations of Delta-BigJoin will not work. Thus, a more proper way is to compare the time increase of each method as the batch size grows.

\subsubsection{Running Environment.}
All methods in our experiments are running on a cluster of $11$ machines, one master and $10$ slaves. The master has $47$GB RAM, two Intel Xeon X5650 CPUs, and one $900$GB HDD. Each slave has $125$GB RAM, two Intel Xeon E5-2630 CPUs, and one $900$GB HDD. We deploy an instance of Apache Hadoop \footnote{http://hadoop.apache.org/} on the cluster. By default we set the number of mappers and reducers to be $200$, each with $4$GB memory space. To run Delta-BigJoin, we also deploy an instance of Timely Dataflow \footnote{https://github.com/frankmcsherry/timely-dataflow} on the cluster, and set the number of processes to be $200$.

\subsubsection{Parameter Settings.}
In the preprocessing step, both SEED and Crystal list all $k$-cliques for $k\leq k_0$ in the data graph, and the user can specify the maximum size $k_0$ of cliques to be listed. In our experiments, we set $k_0$ to be $3$, which is enough for all the $5$ patterns. For DDSL, $m$ is set to be equal to the number of mappers.

\subsubsection{Evaluation Metrics.}
To compare the efficiency of each method, we use the wall-clock time to evaluate the cost of each method. Specifically, we count the elapsed time from submitting a job until it finishes. To show the space cost of NP storage, we use the file size in megabytes as the metric.

\subsection{Experiments on Static Graphs}
We divide the experiments on static graphs into two parts. The first part is to evaluate the computation and space cost of preprocessing and storing the data graph. The second part is to evaluate the cost of listing all matches of a given pattern.

\subsubsection{Preprocessing Costs.}
We first study the cost of constructing the NP storage. In Fig. \ref{fig:exp_pre_time}, we compare the preprocessing time of DDSL with SEED and Crystal. In this step, SEED and Crystal list all $3$-cliques in the data graph, while DDSL only partition edges to each reducer, and write each part of $\Phi(d)$ as a whole in to disks. Compared with constructing $\Phi(d)$, the I/O cost for searching all $3$-cliques is much larger, and thus for all $4$ datasets, DDSL outperforms SEED and Crystal up to $5$ times.

\subsubsection{The Space Cost of NP Storage.}
\begin{figure}[ht]
\centering
\subfloat[Preprocessing time.]{
    \includegraphics[width=.45\linewidth, keepaspectratio]{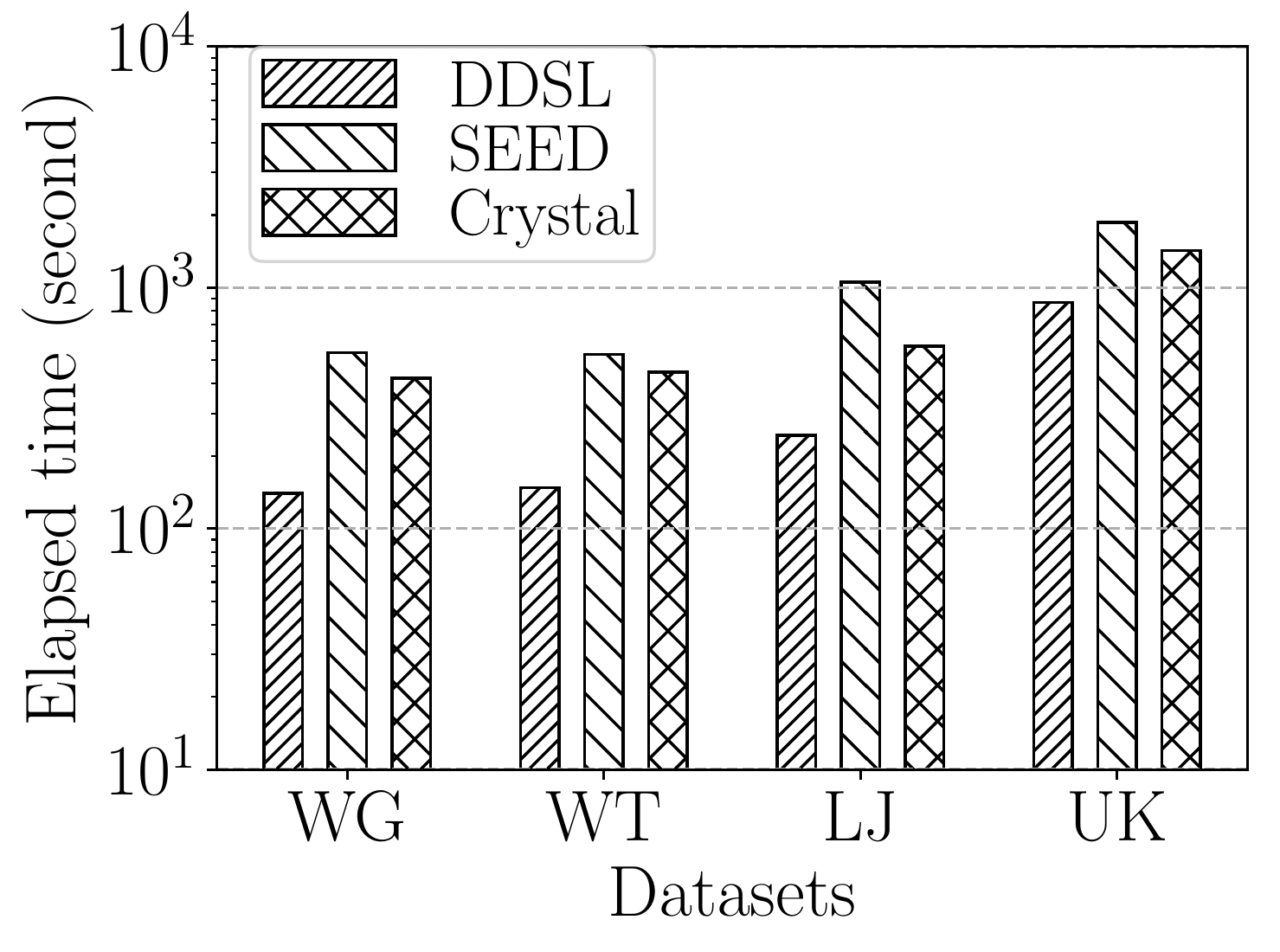}
    \label{fig:exp_pre_time}
}%
\subfloat[Space cost.]{
    \includegraphics[width=.45\linewidth, keepaspectratio]{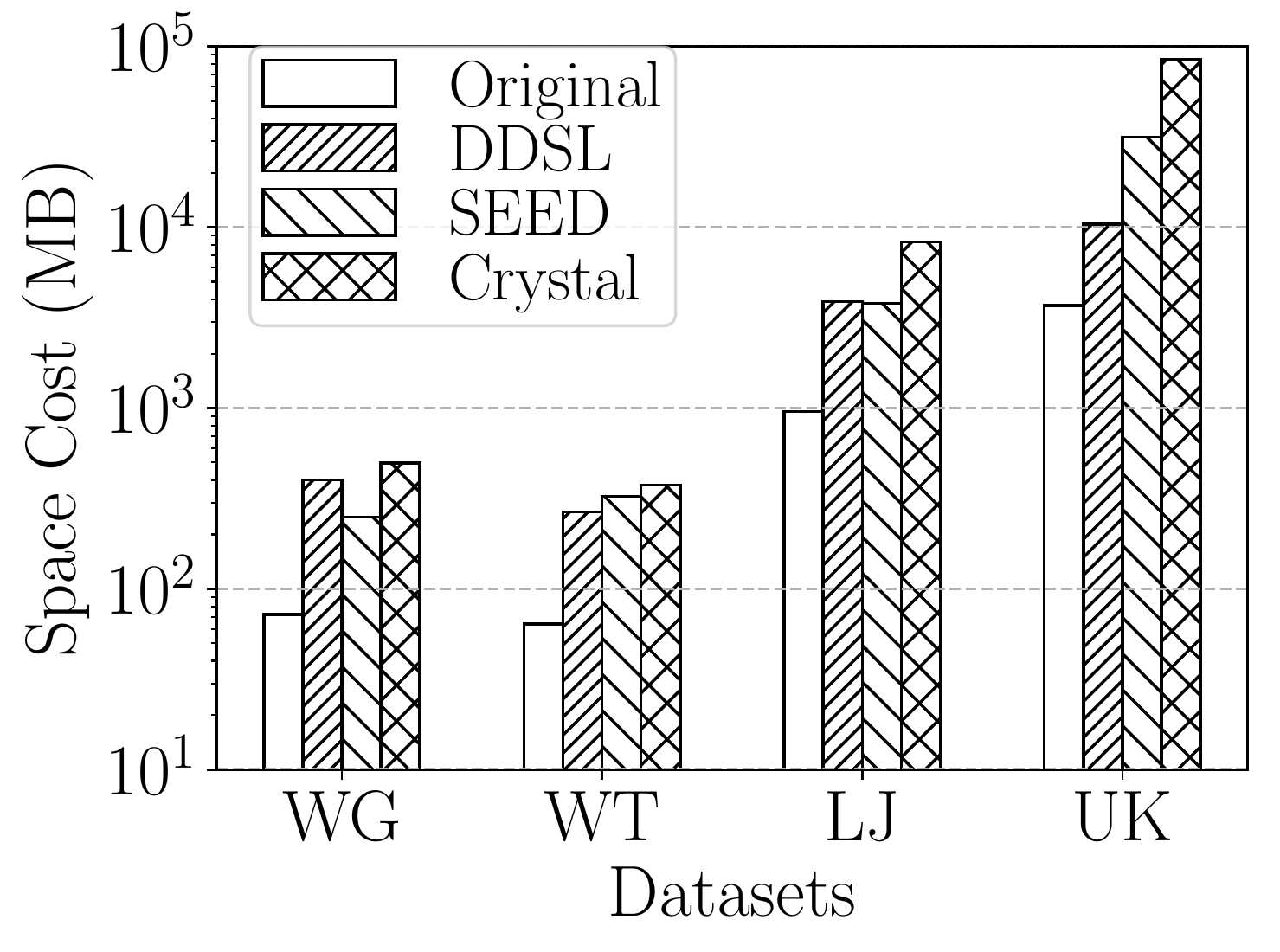}
    \label{fig:exp_pre_space}
}
\caption{Preprocessing cost.}
\label{fig:exp_pre}
\end{figure}

Besides the construction time, we also compares the space cost of each method's underlying storage. Specifically, DDSL uses the NP storage, SEED uses the compact SCP storage, and Crystal stores all $3$-cliques. For each method, the total file size across the cloud is shown in Fig. \ref{fig:exp_pre_space}. Compared to the original graph, the NP storage takes at most $4.6$ times of extra space, while SEED and Crystal can take more than $7$ and $10$ times of extra space, respectively. The major reason is that, both SEED and Crystal break the graph into small structures, resulting in many duplicated edges stored in different parts. On the other hand, DDSL breaks the graph into less parts, so it duplicates less edges across the partitions.

\subsubsection{Overall Performance on Static Graphs.}
\begin{figure*}
\centering
\subfloat[Pattern:$q_1$]{
    \includegraphics[width=.18\linewidth, keepaspectratio]{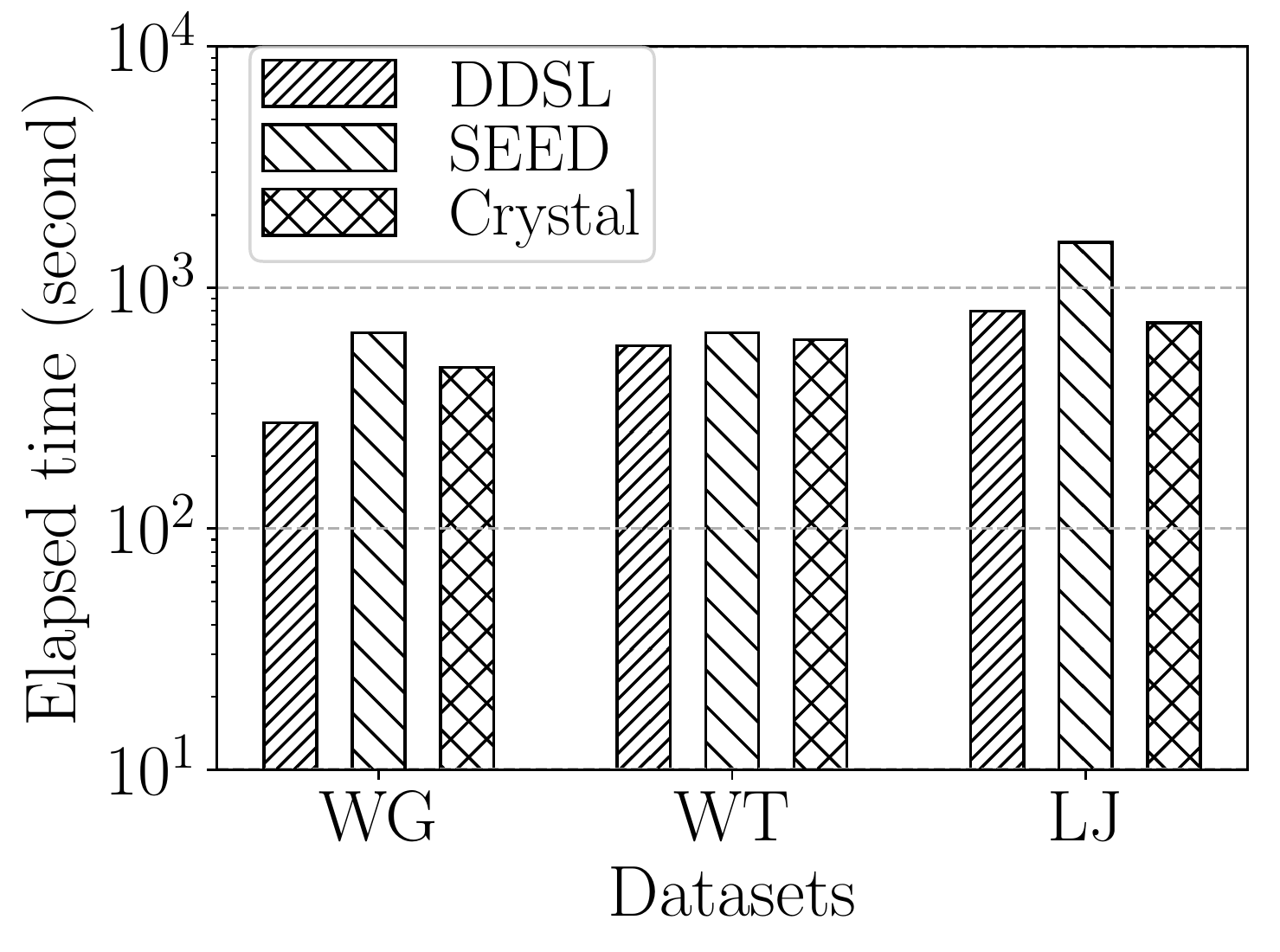}
    \label{subfig:exp_static_q1}
}%
\subfloat[Pattern:$q_2$]{
    \includegraphics[width=.18\linewidth, keepaspectratio]{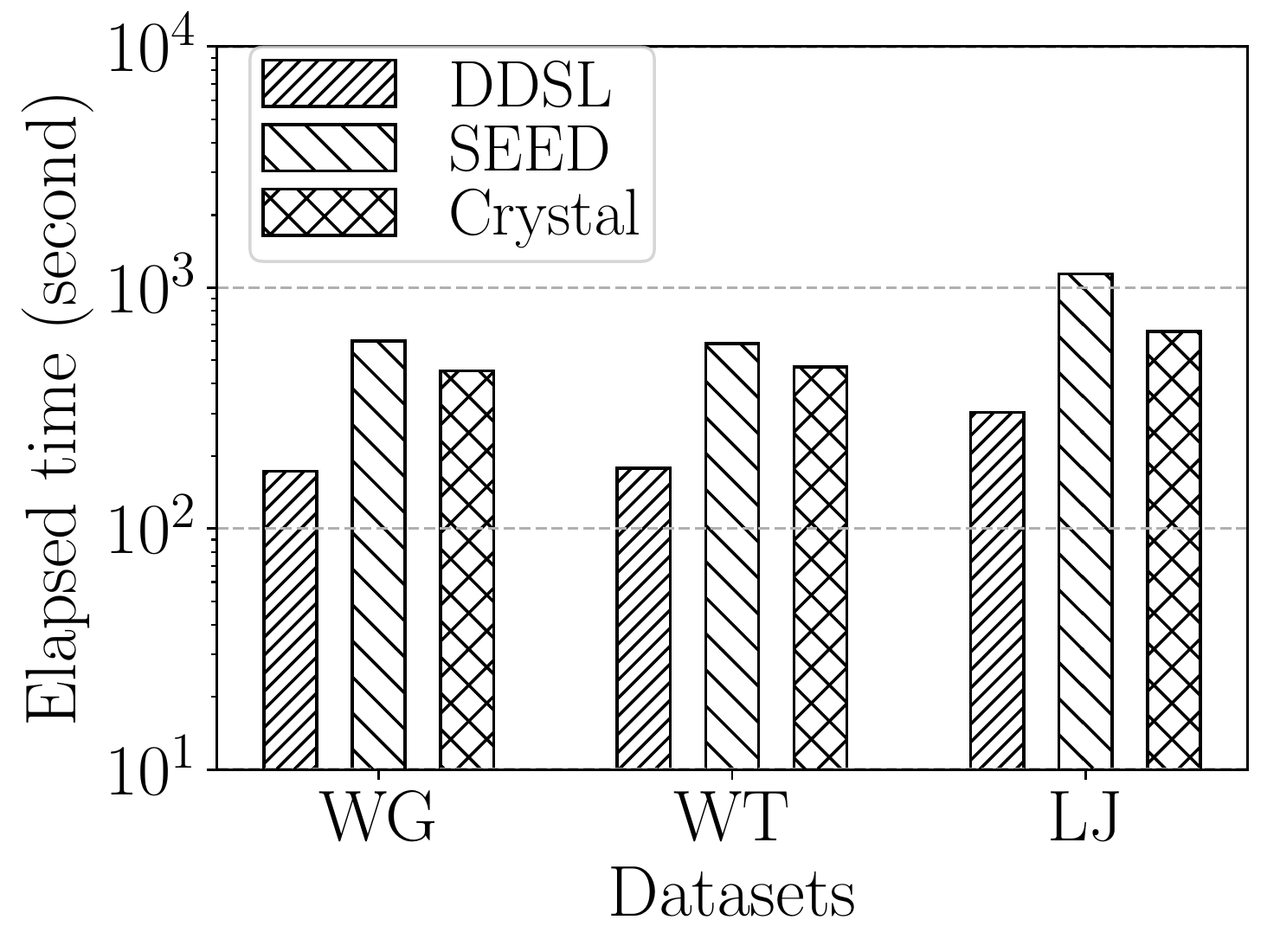}
    \label{subfig:exp_static_q2}
}%
\subfloat[Pattern:$q_3$]{
    \includegraphics[width=.18\linewidth, keepaspectratio]{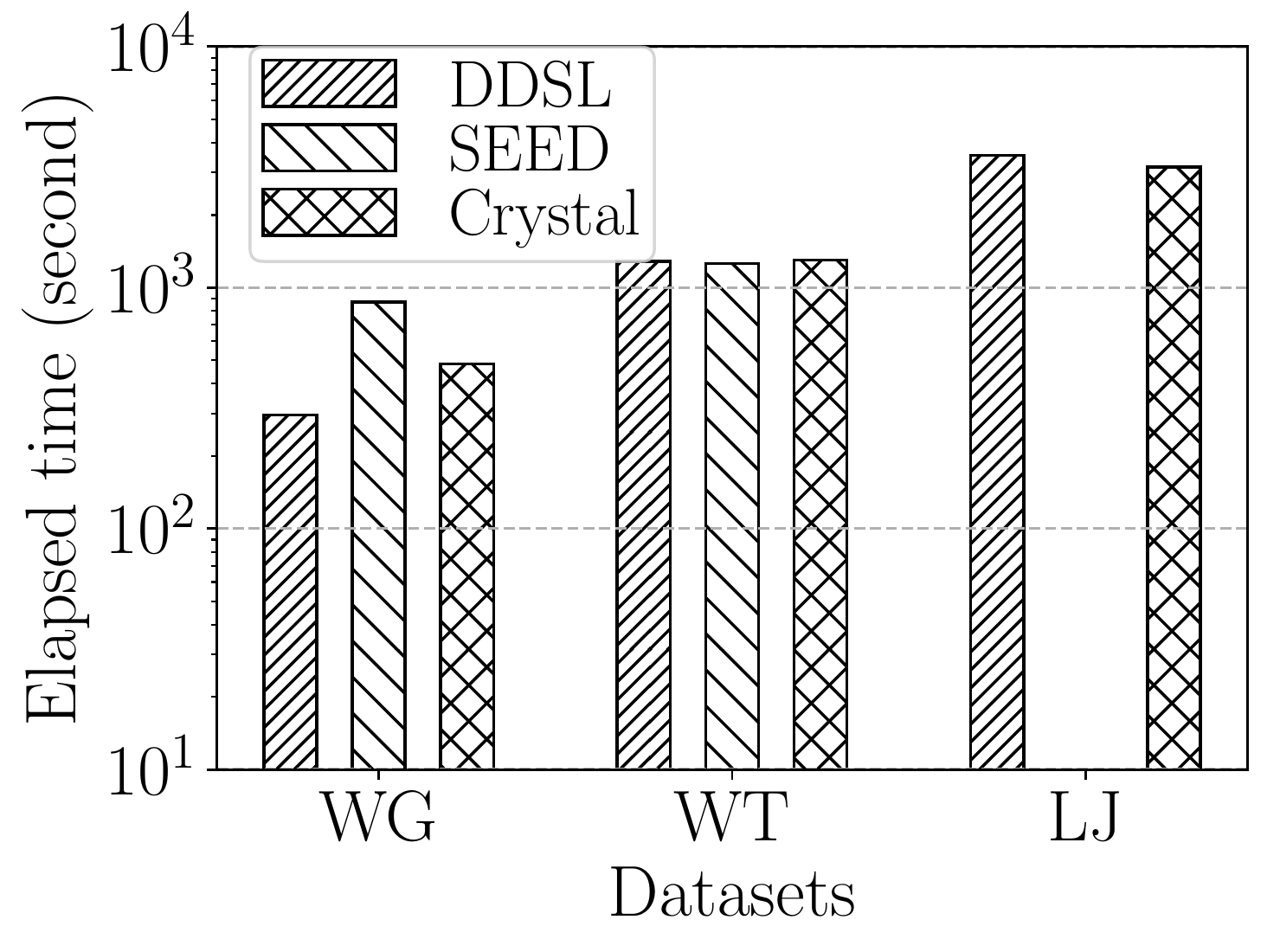}
    \label{subfig:exp_static_q3}
}%
\subfloat[Pattern:$q_4$]{
    \includegraphics[width=.18\linewidth, keepaspectratio]{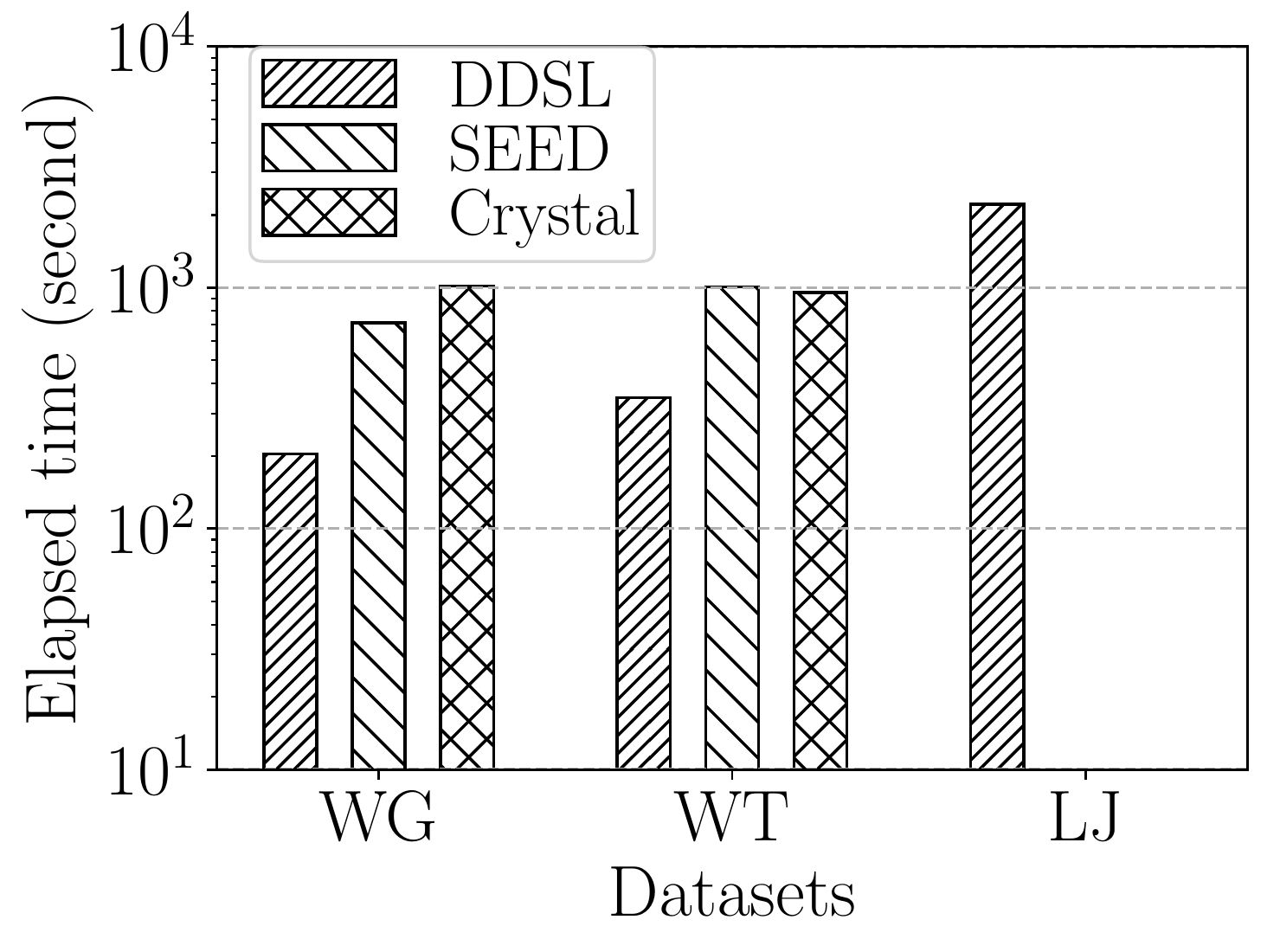}
    \label{subfig:exp_static_q4}
}%
\subfloat[Pattern:$q_5$]{
    \includegraphics[width=.18\linewidth, keepaspectratio]{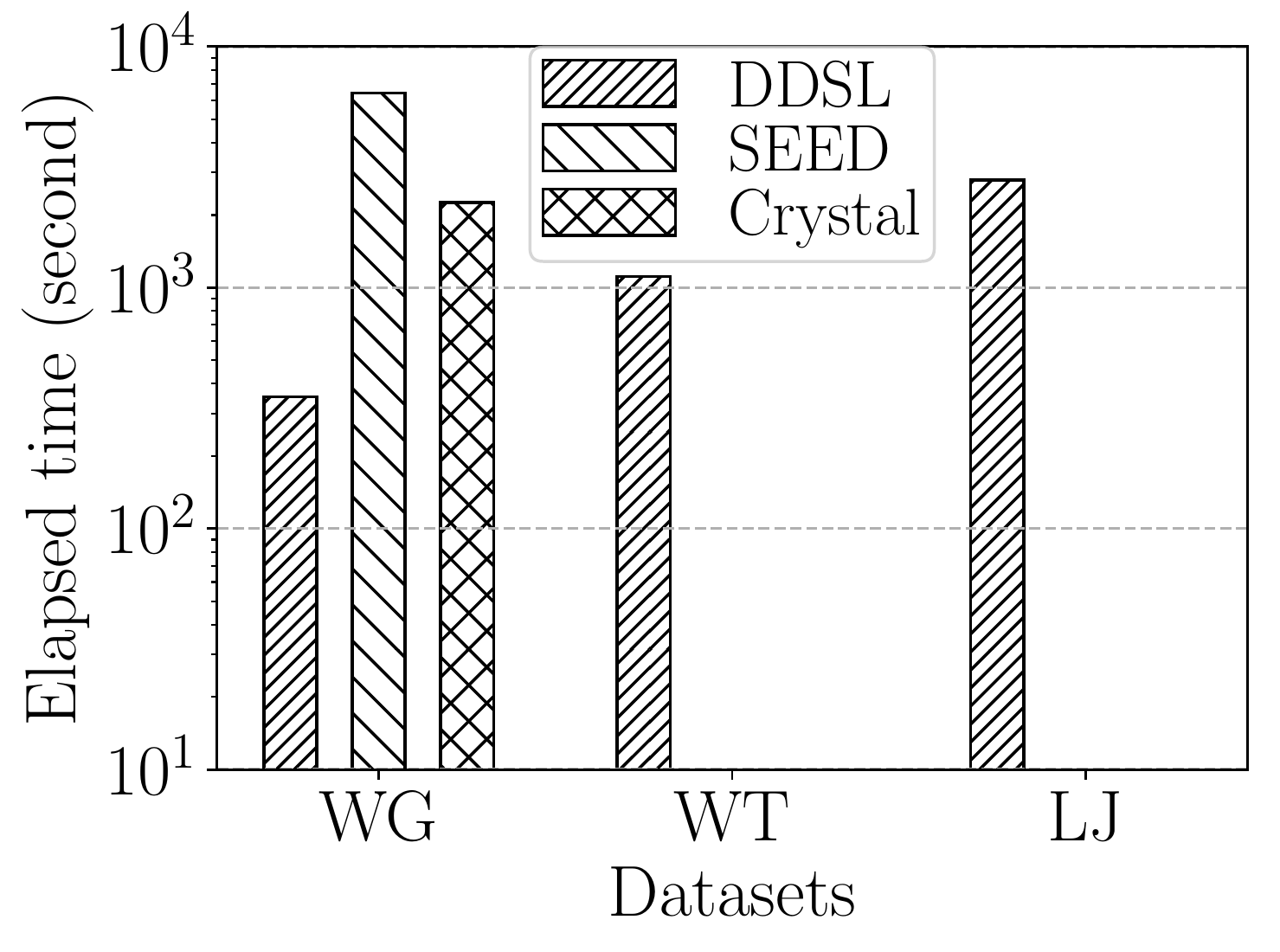}
    \label{subfig:exp_static_q5}
}%
\caption{Performance on static graphs: vary pattern.}
\label{fig:exp_static_q}
\end{figure*}

In this part we study the cost of subgraph listing for each method on static graphs. It is notable that the preprocessing is not independent from the listing process. For example, In SEED and Crystal, one can set $k_0$ to be $4$ in stead of $3$, so that all $4$-cliques are listed. With this additional data, the subgraph listing process of some patterns might be faster, however, listing $4$-cliques incurs more I/O cost, and thus the preprocessing takes longer time. Therefore, we use the overall elapsed time as the measurement to better illustrate the overall performance of each method. In Fig. \ref{fig:exp_static_q} we show the running time of each method on all pattern graphs. The missing bars mean that the running time is larger than $10^4$ seconds. In general, DDSL outperforms other two methods in most of the situations. It is slightly slower than Crystal only for processing $q_1$ and $q_3$ on LJ. For these two patterns, all three methods performs similar joins while listing the matches, and thus DDSL does not have notable advantages over other two methods.
For other three patterns, DDSL has the best performance on all datasets. Specifically, the gap between the running time of DDSL and other two methods are larger when processing more complex patterns. The reason is that, for complex patterns, both SEED and Crystal need to perform several join operations during the process, while DDSL can directly list the matches without join. This result confirms our motivation and the power of supporting R1 units.

\subsection{Experiments on Dynamic Graphs}
Experiments in this part also contain two parts. The first part is to evaluate the cost of updating the NP storage, and the second part is to evaluate the cost of updating the matches of a pattern. For both two parts, we generate the update batch by randomly picking edges. Supposing the batch size is $b$, we randomly picking $b/2$ edges in $d$ to delete, and then randomly generate $b/2$ edges that do not exist in $d$ to insert.

\subsubsection{Cost of Updating the NP Storage}
In this part, we enumerate each batch size in $\{10^2, 10^3, 10^4, 10^5\}$, and count the time of updating the NP storage on each dataset. As Fig. \ref{fig:exp_dynamic_np} shows, DDSL can update the NP storage very efficiently. In comparison to Fig. \ref{fig:exp_pre}, the updating time is less than the construction time even for the largest batch size. We also noticed that for each dataset, the updating time remains nearly unchanged as the batch size grows, which means the message size during the updating also grows slowly.

\subsubsection{Cost of Updating the Result}
To evaluate the performance of updating the match set, we vary the batch size in $\{10^2, 10^3, 10^4, 10^5\}$ on LJ, and count the overall elapsed time of DDSL and Delta-BigJoin. For DDSL, we also include the time for updating of the NP storage, to illustrate its overall performance. We use $inf$ to indicate either the running time is too long, or the memory usage exceeds our capacity.

\begin{figure*}[htbp]
\centering
\subfloat[Updating the NP storage.]{
    \includegraphics[width=.18\linewidth, keepaspectratio]{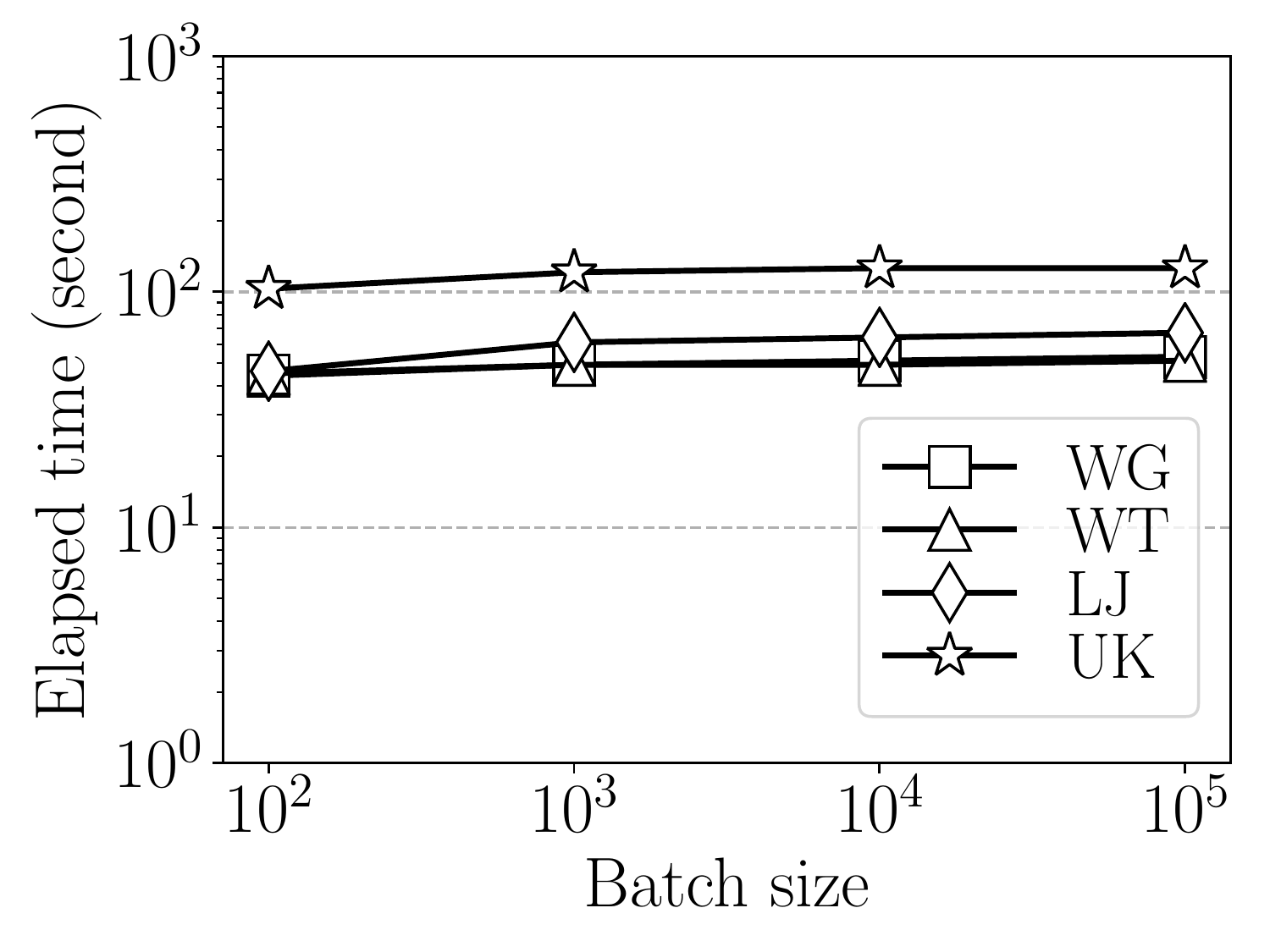}
    \label{fig:exp_dynamic_np}
}%
\subfloat[Updating Matches:$q_1$]{
    \includegraphics[width=.18\linewidth, keepaspectratio]{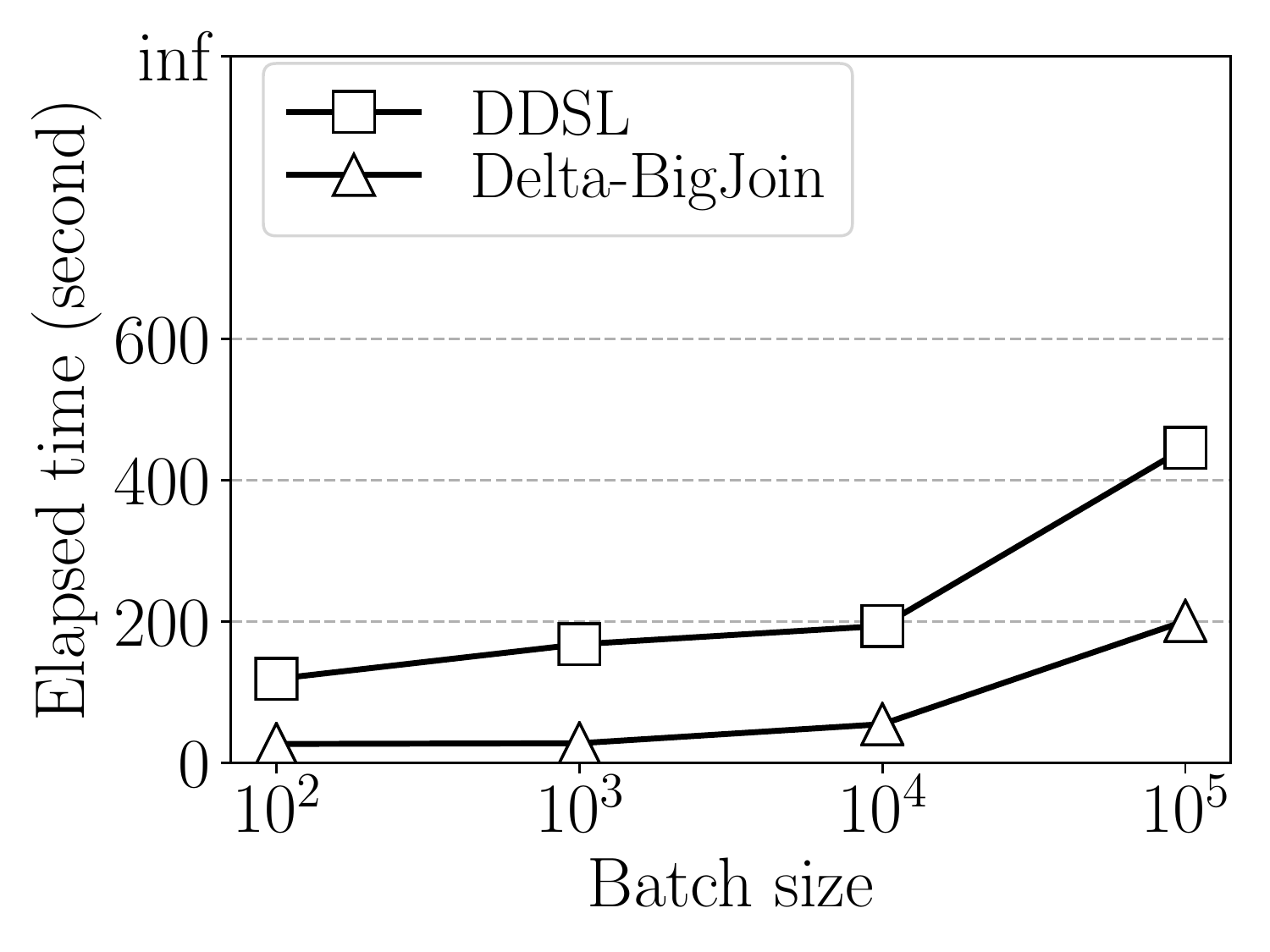}
    \label{subfig:exp_dynamic_q1}
}%
\subfloat[Updating Matches:$q_2$]{
    \includegraphics[width=.18\linewidth, keepaspectratio]{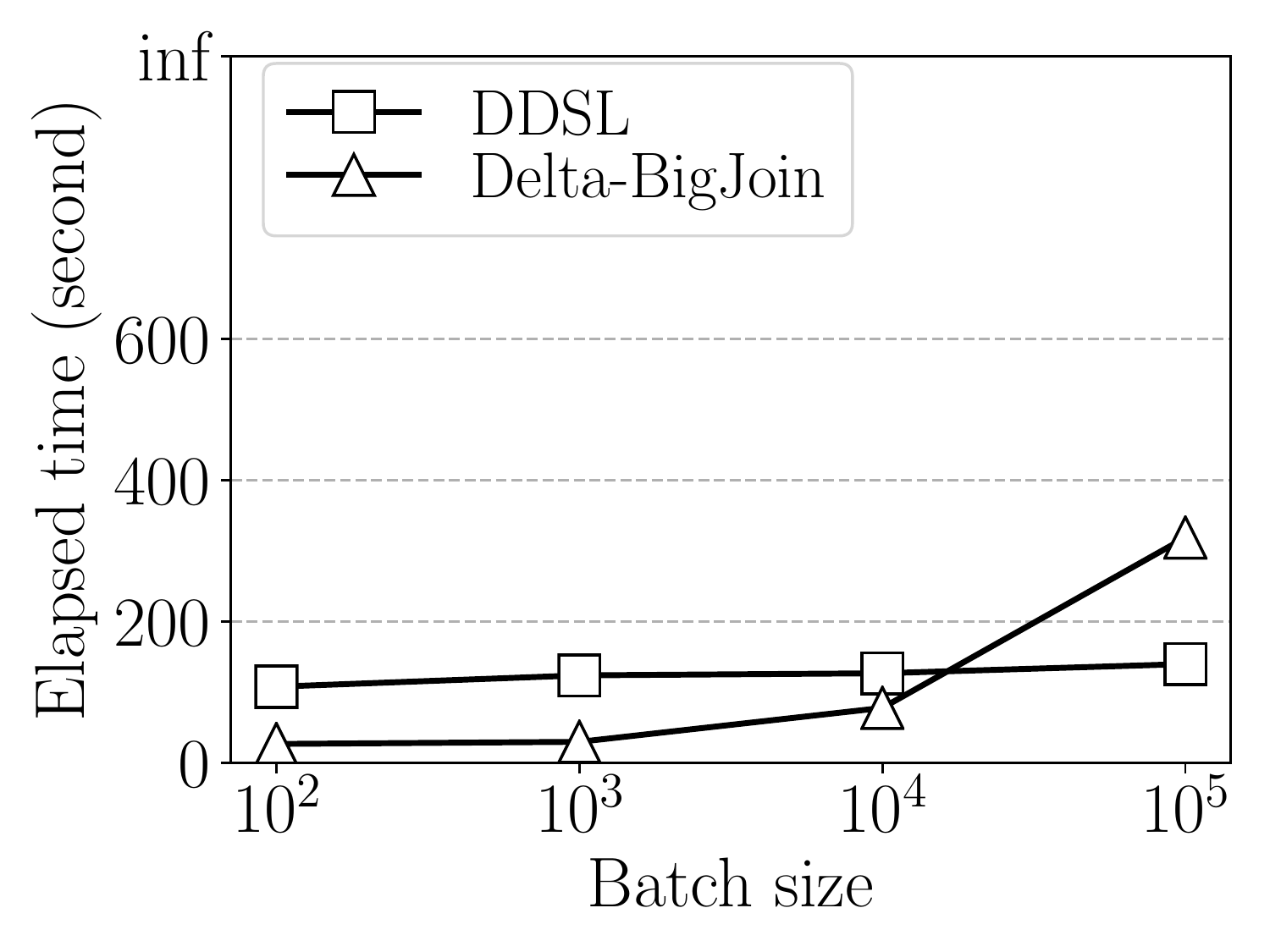}
    \label{subfig:exp_dynamic_q2}
}%
\subfloat[Updating Matches:$q_3$]{
    \includegraphics[width=.18\linewidth, keepaspectratio]{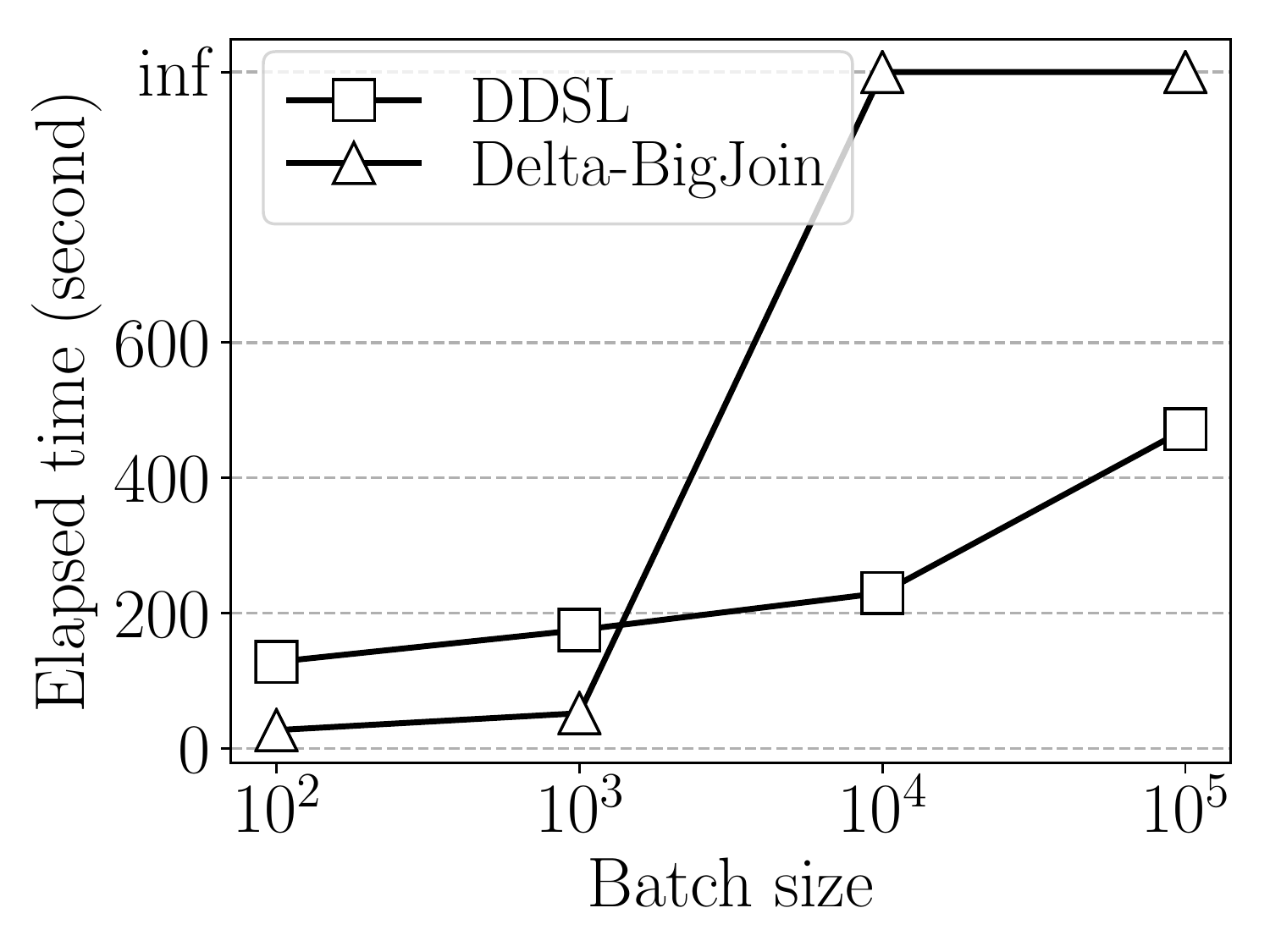}
    \label{subfig:exp_dynamic_q3}
}%
\subfloat[Updating Matches:$q_4$]{
    \includegraphics[width=.18\linewidth, keepaspectratio]{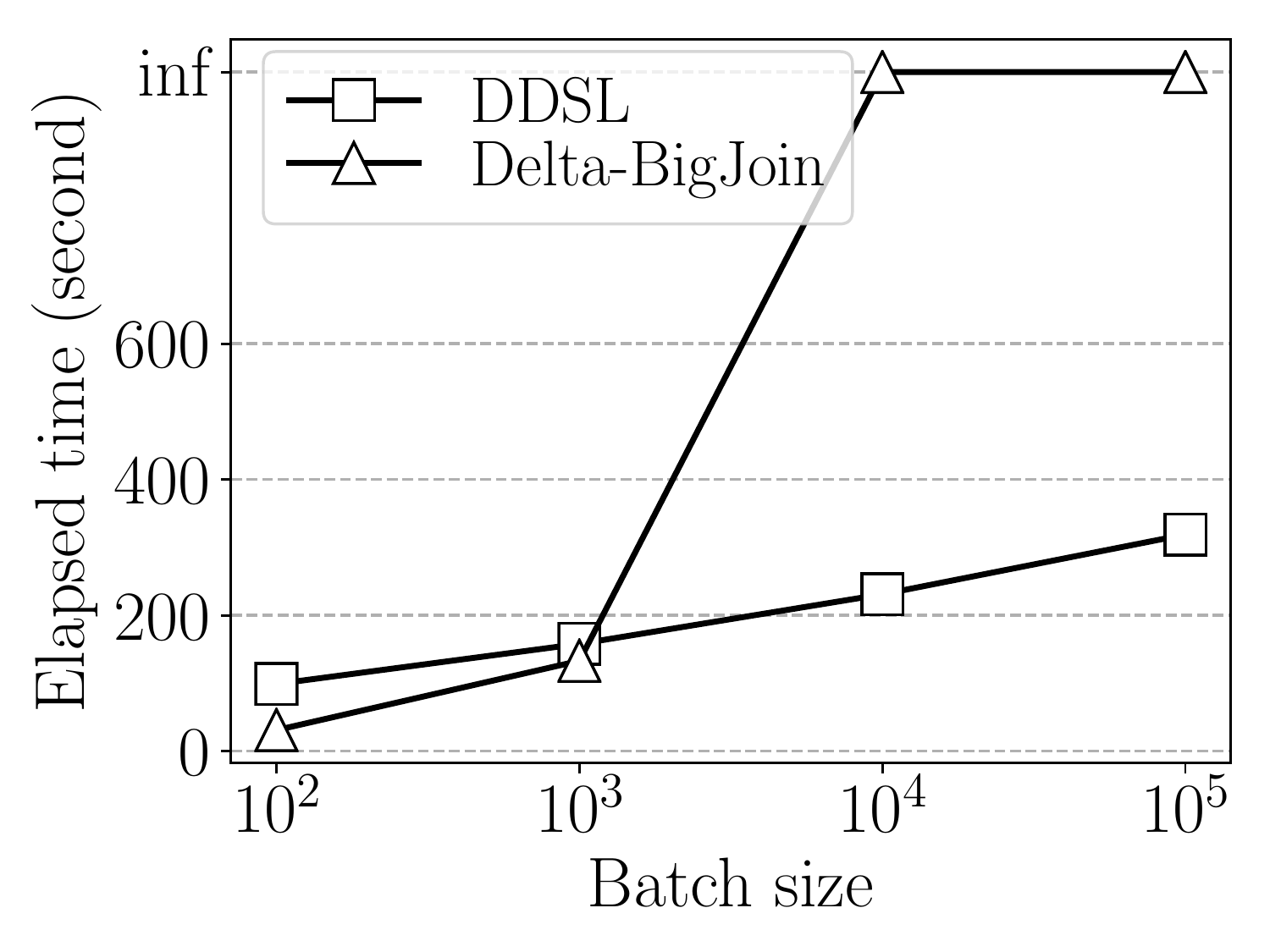}
    \label{subfig:exp_dynamic_q4}
}%
\caption{Performance of incremental updating.}
\label{fig:exp_dynamic}
\end{figure*}

As Fig. \ref{subfig:exp_dynamic_q1} to Fig. \ref{subfig:exp_dynamic_q4} show, except for $q_2$, the increasing of running time of Delta-BigJoin is much faster than that of DDSL. A possible reason is that, Delta-BigJoin maintains several delta queries at the same time, and as the batch size increases, maintaining those queries causes high pressure to the main memory as well as the communication. Actually, for the smallest pattern $q_2$, Delta-BigJoin performs similar to DDSL, which confirms this explanation. On the other hand, DDSL performs stabler. The updating time of all $4$ patterns remains to be low. Actually, comparing the updating time with the running time on LJ in Fig. \ref{fig:exp_static_q}, the updating time for every pattern is much smaller than that on static graphs even for the $10^5$ batch size.

One may also notice that the updating time of $q_1$ and $q_3$ grows faster than that of $q_2$ and $q_4$. The major reason is that, to list the matches of pattern $q_1$ and $q_3$, we need to perform a join operation, while for $q_2$ and $q_4$ we directly list the matches. This, again, shows that reducing the join operations can often reduce the total cost.

\section{Related Work}
\label{sec:related}
\textbf{Centralized Subgraph Listing.} Subgraph listing on a single machine has been extensively studied in the literature. Most of the solutions like VF2 \cite{cordella2004sub}, QuickSI \cite{shang2008taming}, and GADDI \cite{zhang2009gaddi} use backtracking and recursing to find the matches. In GraphQL \cite{he2008graphs} and SPath \cite{zhao2010graph}, neighbors' labels of a vertex are used to filtering invalid matches at early steps. Han et al. \cite{han2013turbo} finds that the matching order can significantly influence the efficiency of the algorithm. More recently, Kim et al. \cite{kim2016dualsim} proposes an I/O-efficient algorithm through a dual approach. 

\textbf{Distributed Subgraph Listing.} Many approaches are proposed recently to solve the subgraph listing problem in a distributed environment. Sun et al. \cite{sun2012efficient} use the Trinity memory cloud to parallelize a join-based algorithm, which employs STwigs as the join unit. Shao et al. \cite{Shao:2014:PSL:2588555.2588557} parallelize the traditional DFS algorithm using Pregel \cite{malewicz2010pregel}. They use several pruning rules as well as the workload-balancing strategy to improve the efficiency. Lai et al. \cite{lai2015scalable,lai2016scalable} investigate the join-based algorithms based on MapReduce. They try to reduce the overall I/O cost by introducing different join units and join trees. Gao et al. \cite{gao2014continuous} achieve approximate subgraph listing through message passing. They convert the query graph into a DAG, and use Giraph \footnote{http://giraph.apache.org/} to pass messages between vertices. Qiao et al. \cite{qiao2017subgraph} propose a framework to compress the matches with the vertex-cover-based-compression, which can further reduce the I/O cost. More recently, Khaled et al. \cite{ammar2018distributed} propose a join algorithm for directed graphs based on a the Timely Dataflow \cite{murray2013naiad} system, which achieves worst-case optimality in terms of computation and communication cost. However, it brings storage pressure to the system by requiring to store the complete data graph on each machine. Also, how to support undirected patterns efficiently in this approach is still an open problem.

\textbf{Subgraph Listing on Dynamic Graphs.} There are only a few works targeting on dynamic graphs. In \cite{gao2014continuous}, matches are identified by passing messages between vertices. When edges are inserted/deleted, the +/- messages are passed through the graph to update vertex status as well as the matching results. This method is efficient since many messages are merged within a vertex, however, this also leads to an inexact result. Fan et al. \cite{fan2017incremental} investigate several problems on dynamic graphs, and propose an exact centralized algorithm for dynamic subgraph enumeration. To handle graph changes, it extracts the union of $d_p$-neighbors for all inserted edges, and applies a centralized static algorithm on this extracted graph. However, as a centralized algorithm, it suffers from the huge I/O cost for the subgraph listing problem on big graphs. For directed graphs, Khaled et al. \cite{ammar2018distributed} propose an algorithm to handle graph changes as streaming data.

\section{Conclusion}
\label{sec:conclusion}
In this paper, we study the problem of subgraph listing on distributed and dynamic graphs. We propose an efficient method, called DDSL, to handle dynamic graphs through two stages: initial calculation and incremental updating. The initial calculation follows a general distributed join framework. In order to reduce the join operations, we use the bounded neighbor-preserved storage mechanism for the data graph, which supports listing the matches of any R1 unit directly. To further reduce the intermediate result size, we incorporate the existing vertex-cover-based compression into this framework. To choose a better join order, we derive a comprehensive cost model, and use a dynamic programming to find the optimal join tree. In the incremental updating stage, we first design an algorithm to update the NP storage, then we propose the novel Nav-join to compute newly-appeared matches. Extensive experiments show that DDSL can handle static subgraph listing with a competitive performance compared with the state-of-the-art distributed methods. Moreover, DDSL can efficiently handle dynamic subgraph listing without computing from scratch. To the best of our knowledge, DDSL is the first approach in the literature that supports unlabeled and undirected pattern matching on dynamic graphs in a distributed environment.




\let\xxx=\bibitem\def\bibitem{\par\vspace{-0.2mm}\xxx} 

\bibliographystyle{IEEEtran}
\bibliography{references}

\end{document}